\documentclass[twocolumn,preprint]{aastex631}

\usepackage{CJK}
\usepackage{apjfonts}
\usepackage{amsmath}
\usepackage{color}

\hypersetup{
   colorlinks,
   linkcolor={blue!88!black!80},
   citecolor={blue!88!black!80},
   urlcolor={blue!88!black!80}}

\shorttitle{Stream collision and delayed disk formation in TDEs}
\shortauthors{Guo et al.}

\turnoffeditone

\begin{document}
\begin{CJK}{UTF8}{gbsn}

\title{Reverberation evidence for Stream Collision and Delayed Disk Formation in Tidal Disruption Events}

\author[0000-0001-8416-7059]{Hengxiao Guo (郭恒潇)}
\affiliation{Shanghai Astronomical Observatory, Chinese Academy of Sciences, 80 Nandan Road, Shanghai 200030, People's Republic of China}
\affiliation{SHAO-XMU Joint Center for Astrophysics, Xiamen, Fujian 361005, People's Republic of China}

\correspondingauthor{Hengxiao Guo, Jingbo Sun, Yan-Fei Jiang}
\email{hengxiaoguo@gmail.com (HXG) \\
sunjingbo@shao.ac.cn (JBS)\\
yjiang@flatironinstitute.org(YFJ)} 

\author[0000-0001-8416-7059]{Jingbo Sun (孙静泊)} 
\affiliation{Shanghai Astronomical Observatory, Chinese Academy of Sciences, 80 Nandan Road, Shanghai 200030, People's Republic of China}
\affiliation{University of Chinese Academy of Sciences, 19A Yuquan Road, 100049, Beijing, People's Republic of China}

\author[0000-0002-7299-4513]{Shuangliang Li}
\affiliation{Shanghai Astronomical Observatory, Chinese Academy of Sciences, 80 Nandan Road, Shanghai 200030, People's Republic of China}

\author[0000-0002-2624-3399]{Yan-Fei Jiang (姜燕飞)}
\affiliation{Center for Computational Astrophysics, Flatiron Institute, New York, NY 10010, USA}

\author[0000-0002-1517-6792]{Tinggui Wang}
\affiliation{CAS Key Laboratory for Research in Galaxies and Cosmology, Department of Astronomy, University of Science and Technology of China, Hefei, 230026, People's Republic of China}
\affiliation{School of Astronomy and Space Sciences, University of Science and Technology of China, Hefei, 230026, People's Republic of China}

\author[0000-0002-0427-520X]{Defu Bu}
\affiliation{Shanghai Astronomical Observatory, Chinese Academy of Sciences, 80 Nandan Road, Shanghai 200030, People's Republic of China}

\author[0000-0002-7152-3621]{Ning Jiang}
\affiliation{CAS Key Laboratory for Research in Galaxies and Cosmology, Department of Astronomy, University of Science and Technology of China, Hefei, 230026, People's Republic of China}
\affiliation{School of Astronomy and Space Sciences, University of Science and Technology of China, Hefei, 230026, People's Republic of China}

\author[0000-0003-3207-5237]{Yanan Wang}
\affiliation{National Astronomical Observatories, Chinese Academy of Sciences, 20A Datun Road, Beijing 100101, People's Republic of China}

\author[0000-0001-6747-8509]{Yuhan Yao}
\affiliation{Miller Institute for Basic Research in Science, 468 Donner Lab, Berkeley, CA 94720, USA}
\affiliation{Department of Astronomy, University of California, Berkeley, CA 94720, USA}

\author[0000-0001-5012-2362]{Rongfeng Shen}
\affiliation{School of Physics and Astronomy, Sun Yat-Sen University, Zhuhai, 519082, People's Republic of China}
\affiliation{CSST Science Center for the Guangdong-Hongkong-Macau Greater Bay Area, Sun Yat-Sen University, Zhuhai, 519082, People's Republic of China}

\author[0000-0002-4455-6946]{Minfeng Gu}
\affiliation{Shanghai Astronomical Observatory, Chinese Academy of Sciences, 80 Nandan Road, Shanghai 200030, People's Republic of China}

\author[0000-0002-0771-2153]{Mouyuan Sun}
\affiliation{Department of Astronomy, Xiamen University, Xiamen, Fujian 361005, People's Republic of China}

\begin{abstract}
When a star passes through the tidal disruption radius of a massive black hole (BH), it can be torn apart by the tidal force of the BH, known as the Tidal Disruption Event (TDE). Since the observed UV/optical luminosity significantly exceeds the predictions of the compact disk model in classical TDE theory, two competing models, stream collision and envelope reprocessing, have been proposed to address this discrepancy. To distinguish between these models, we investigate the continuum reverberation behaviors for $\sim$30 TDEs with high-quality multi-band light curves. We found that over half of them exhibit a positive lag by a few days in UV/optical bands, indicating that their inferred sizes are significantly larger than the envelope sizes in reprocessing. Moreover, X-ray emissions not only significantly delay relative to the primary UV/optical peak but also lag behind the rebrightening bump by up to several tens of days, completely different from the X-ray illumination reprocessing. Additionally, the anti-correlated UV-optical continuum in ASASSN-15lh further disfavors the reprocessing scenario. In contrast, the model of stream collisions, combined with delayed accretion disk formation, can provide a unified explanation for the diverse TDE observations, e.g., the optical/X-ray population, the frequently observed rebrightening bump. This model describes a unification scheme wherein the UV/optical emission originates from stream collisions during the early-stage of TDE evolution and gradually transitions to being dominated by accretion disk with detectable X-ray emission in the late stage. After transitioning to a quiescent state, recurrent flares may be observed in some cases, possibly related to repeating partial TDEs. 

\end{abstract}


\section{Introduction} 
\label{sec:intro}
According to the classical theory of Tidal Disruption Event \citep[TDE,][]{Rees88,Evans89}, a star will be torn apart when it crosses into the tidal radius of a massive black hole (BH). Around half of the stellar debris would escape at high speed following a parabolic trajectory while the remaining material will fall back to the BH and form a nascent accretion disk. The circularization of the disk formation could produce a luminous electromagnetic flare and then gradually decline following a power-law profile \citep{Phinney89}, regulated by the mass fallback rate. The classical TDE theory predicts that a compact accretion disk forms around the tidal radius, which is approximately $10^{13}$ cm for the disruption of a solar-like star by a BH with a mass of $M_{\rm BH} = 10^{6}$~$M_{\odot}$. The thermal emission at this radius is expected to peak at the soft X-ray and exhibits very low UV/optical luminosity, typically less than $10^{42}$ erg~s$^{-1}$. However, with the growing number of optical-selected TDEs discovered by modern time-domain surveys, such as the All-Sky Automated Survey for Supernovae \citep[ASASSN,][]{Shappee14} and Zwicky Transient Facility \citep[ZTF,][]{Bellm19}, the community found that the observed UV/optical luminosity is significantly higher than the prediction of a circularized debris disk. In other words, the UV/optical emitting radius, as derived from the single-temperature blackbody assumption, is substantially larger than the size predicted for a classical accretion disk. This suggests that the UV/optical emission of TDEs may not originate from a classical accretion disk \citep[see a recent review, ][]{Gezari21}.

To resolve the tension between observations and theory, two competing models were proposed: the first is the stream collision (or shock-disk scenario), which predicts that the trajectory and morphology of the fallback debris stream will be affected by the relativistic effects (i.e., relativistic apsidal precession, nozzle shock, and spin-induced nodal precession) at pericenter \citep{Bonnerot21}, leading to a collision between the subsequent outward stream and previous infalling stream. This collision-induced shock thus produces the observed UV/optical flare \citep{Piran15,Shiokawa15,Jiang16,Bonnerot21,Huang23,Ryu23}, and the accretion disk forms as the gas falls back to the BH, giving rise to multi-wavelength emission. The second is the envelope reprocessing \citep{Loeb97,Metzger16,Roth16,Dai18,Parkinson22,Thomsen22,Metzger22}. As the massive BH accretes the fallback gas at a super-Eddington ratio after a fast circularization, the gas forms a radiatively inefficient slim disk \citep{Abramowicz88} with strong outflows. These high-velocity outflows take gas to a larger radius, forming a quasi-spherical and optically thick envelope (or winds/outflows) that reprocesses the X-ray/EUV emission from the inner disk to the observed UV/optical emission. Due to the high opacity of the envelope around the accretion disk, the X-ray is thought to be temporarily obscured by ubiquitous outflows during the peak of the UV/optical flare and gradually becomes transparent as the outflow mass rate decreases. While the unexpectedly large UV/optical emitting radius can be explained by both scenarios, continuum reverberation mapping (CRM) may offer critical clues to distinguish between them. This will yield pivotal insights into the UV/optical emission origins of TDEs, thereby clarifying the overall picture of TDEs and serving as a foundation for further in-depth investigations.

In this work, we first introduce our sample selection and lag measurements in \S \ref{sec:sample}. Next, we present the CRM results in TDEs and discuss their implications in \S \ref{sec:results} \& \ref{sec:theory}. Finally, we propose a unification model for TDE evolution, explain how it aligns with current observations in \S \ref{sec:unificaiton}, and draw our conclusions in \S \ref{sec:conclusion}. Throughout this paper, we use the $\Lambda$CDM cosmology, with $H_{\rm 0}$ = 70.0 $\rm km~ s^{-1}~ Mpc^{-1}$, $\Omega_{\rm M}$ = 0.3 and $\Omega_{\rm \Lambda}$ = 0.7.

\section{Target selection and lag measurements}\label{sec:sample}
\subsection{Target Selection}
To conduct CRM for lag measurements and thereby elucidate the origin of their UV/optical emission, we require high-quality optical TDE light curves that meet specific cadence requirements. The ZTF, an optical time-domain survey utilizing the 1.2 m (48 inch) Schmidt telescope at Palomar Observatory, has a wide field of view (47 deg$^2$) and a high-cadence ($\sim$2$-$3 days) in the $gri$ bands. Therefore, we primarily focus on TDEs detected by ZTF for two reasons: firstly, ZTF is responsible for discovering the majority of TDEs, providing uniform data quality; secondly, ZTF's multi-band light curve sampling meets the requirements necessary for accurate lag measurement.

For the ZTF TDEs, the complete dataset consists of two subsets: 30 objects reported by \citet{vanVelzen21b} and \citet{Hammerstein23}, alongside 33 objects independently selected by \citet{Yao23}. Among these, 13 sources overlap between the two groups, resulting in a total of 50 unique TDEs with redshifts up to 0.5. Despite differences in datasets and selection criteria, the primary identification steps are similar: targets are initially detected through real-time alerts and classified as nuclear transients. Subsequently, photometric and spectroscopic properties are employed to thoroughly exclude impostors such as stellar and Active Galactic Nucleus (AGN) variability, resulting in 50 spectroscopically confirmed TDEs. For more details, we refer the reader to the original references.

\begin{figure*}
\centering
\includegraphics[width=0.9\linewidth]{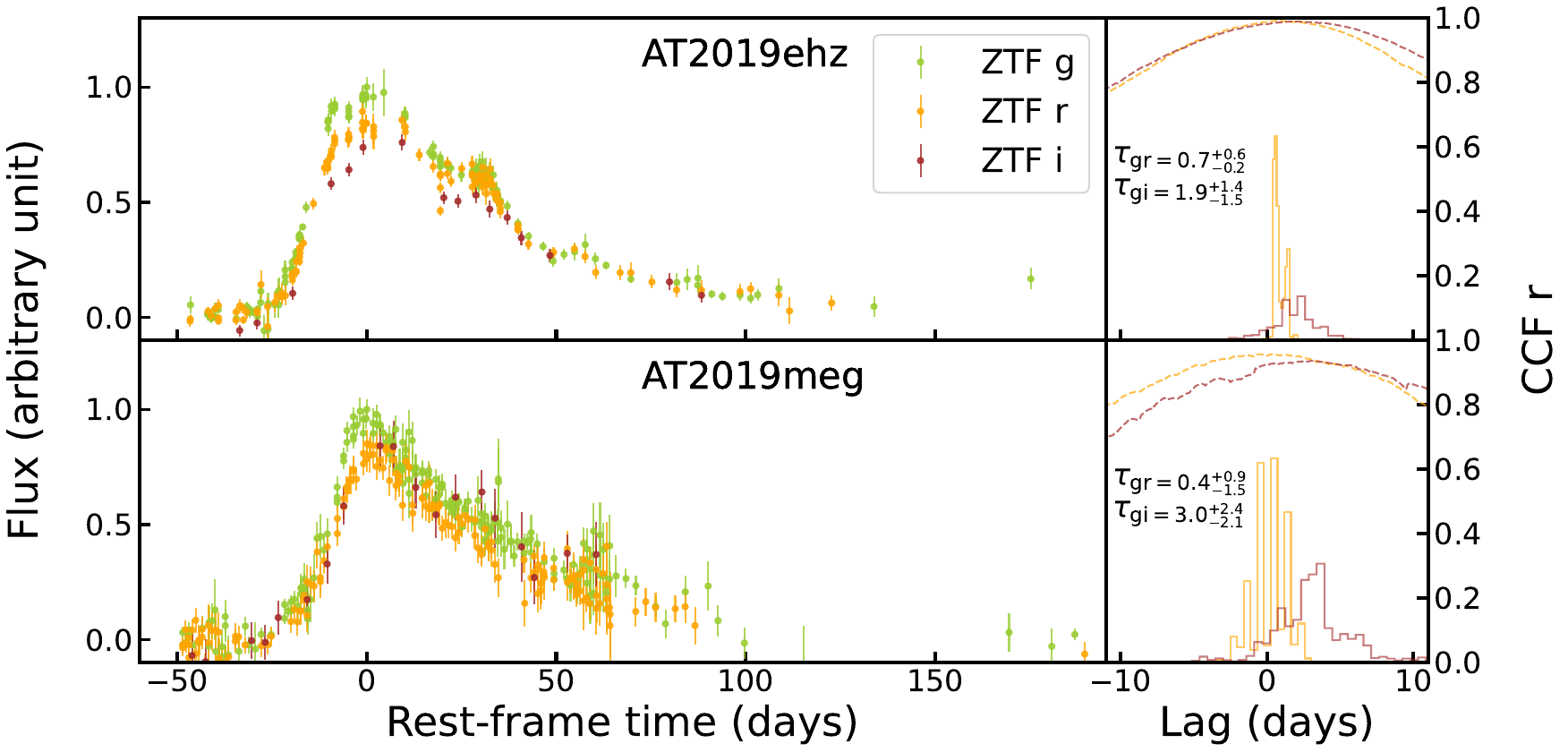}
\caption{Two examples of ZTF TDE light curves (left panels) with ICCF lag measurements (right panels). All the light curves are host subtracted. It is clear that the flare peaks of the longer wavelengths are delayed relative to the $g$ band maximum light, although the $i$-band cadence is much lower than that of $gr$ bands. The CCF curves (dashed lines) and lag posteriors are plotted in the right panel.}
\label{fig:example}
\end{figure*}

For lag measurement purpose, we further excluded 18 out of 50 objects from our analysis due to incomplete coverage caused by seasonal gaps at the peak of the UV/optical flare, which is essential for cross-correlation analysis. Our final sample comprises 32 TDEs, including 31 with available $gr$-band light curves and 15 with $gi$-band light curves. Two examples are illustrated in Figure \ref{fig:example}. Additionally, three TDEs (AT 2019azh, AT 2020zso, and AT 2019ehz) have sufficient UV photometries for effective lag detection. 

%(see Table S\ref{tab:ztf_sample}). 

%In addition, our study also highlights three notable TDEs (AT 2019qiz, ASASSN-15lh, and AT 2018fyk) and several associated TDEs (AT 2020ocn, AT 2019avd, AT 2020vdq, and AT 2022dbl)

In addition, our study also highlights three notable TDEs and several associated TDEs:  
\begin{itemize}
    \item {\bf At 2019qiz} is a very close TDE ($z = 0.015$) discovered by ZTF, and well monitored by \citet{Nicholl20} and \citet{Hung21}. Its BH mass is log (M$_{\rm BH}$/M$_{\odot}$) = 6.16 $\pm$ 0.43 based on the $M_{\rm BH}-\sigma$ relation. The unprecedented quality (high cadence) of multi-band light curves is the primary reason we choose this target, as it is extremely important for lag measurements (see \S \ref{sec:first_peak} and Figure \ref{fig:2019qiz}). It is classified as a Bowen TDE according to the spectral classification \citep{vanVelzen21b}, and its late-time spectrum, taken 157 days after the UV/optical peak, suggests that the host galaxy likely harbors a weak AGN \citep{Nicholl20}.
    
    %The velocity dispersion of $\sigma$ is 69.7 $\pm$ 2.3 \kms\ obtained from Very Large Telescope/X-shooter spectrum, indicating a BH mass of log (M$_{\rm BH}$/M$_{\odot}$) = 6.16 $\pm$ 0.43 based on the $M_{\rm BH}-\sigma$ relation \citep{Xiao11}. 
    
    %The delayed broad \ha\ line relative to continuum and narrow coronal lines have also been reported \citep{Nicholl20, Short23}. 
    %Two-epoch polarization observations reveal very low-level continuum polarization ($<$ 1\%) at the initial stage (peak and peak$+$29 days) \citep{Patra22}.

    \item {\bf ASASSN-15lh} is a luminous transient ($z = 0.233$) discovered by ASAS-SN. It was initially classified as a superluminous supernova \citep{Dong16,Godoy-Rivera17}, but subsequent evidence (e.g., temperature evolution) suggests it is a TDE originating from a fast-spinning Kerr BH with a mass of $\sim 10^{8.5} M_{\odot}$ \citep{Leloudas16,Margutti17,Kruhler18,Gezari21}. We incorporate this target due to a distinguishing feature (see \S \ref{sec:second_peak} and Figure \ref{fig:15lh}): the significant rebrightening (RB) bump observed in the UV light curves, also see two similar cases AT 2019avd \citep{Wang23} and AT 2020ocn \citep{Hammerstein23}.
    
    \item {\bf AT 2018fyk} is a both UV/optical and X-ray bright TDE at $z$ = 0.059, first discovered by ASAS-SN. The BH mass is estimated to be log (M$_{\rm BH}/M_{\odot}$) = 7.7 $\pm$ 0.4, based on the M$_{\rm BH}-\sigma$ relation. Its UV and X-ray follow-up lasts over 1300 days from a high accretion state to a quiescent state (see \S \ref{sec:recurrent} and Figure \ref{fig:fyk}), allowing us to explore the correlations of X-ray/UV/optical in different stages of this TDE \citep{Wevers19a,Wevers20,Wevers21}. Remarkably, it displays a recurrent flare a few years later, marking it as a potential repeating partial TDE (rpTDE), also see two similar cases AT 2020vdq \citep{Somalwar23} and AT 2022dbl \citep{Lin24} in \S \ref{sec:recurrent}. 
\end{itemize}

%This event bears similarities to ASASSN-15lh, displaying a distinct rebrightening plateau in the UV/optical spectrum, accompanied by high-cadence X-ray monitoring, although the peak of the UV/optical flare is not captured.

%AT 2018qiz, ASASSN15lh, and AT 2018fyk. AT2018qiz is distinguished by its extensive data sampling and detailed pre-peak information. ASASSN15lh is characterized by a unique UV rebrightening bump. Meanwhile, AT2018fyk is a potentially repeating partial TDE, which has been extensively monitored in both UV and X-ray over a long-term baseline. For more details, see \S \ref{sec:RM}.

\subsection{Lag measurements}
\subsubsection{Cross-correlation analysis}\label{sec:ccf}
To measure the continuum lag, we adopt the most commonly used technique, interpolated cross-correlation function \citep[ICCF,][]{Peterson98} with the Python package {\tt PyCCF} \citep{Sun18} following \citet{Guo22b}. 
%It simply determines the lag at which the correlation between the two light curves is maximized, independent of any assumptions about whether the two light curves are related to each other. 

All the host-subtracted ZTF light curves are moved to the rest frame time. The lag search range is usually set to $\pm$30 days in the rest frame with a 0.1-day grid spacing, except for some X-ray light curves with longer delays (search range slightly extended). The correlation coefficient ($r_{\rm max}$) describes the strength and direction of the correlation. ICCF utilizes the traditional flux randomization/random subset sampling (FR/RSS) procedure \citep{White94,Peterson98} to obtain the lags and uncertainties with 10000 iterations using the Monte Carlo method. The centroid lags are generally adopted as they are more representative of responsivity radius of the variability signal \citep[e.g.,][]{Gaskell86}. For sanity check, we also confirmed that all centroid lags are broadly consistent with peak lags (see \S \ref{sec:first_peak}). To best demonstrate the inter-band lags, we usually select the bluest or reddest band as a reference band depending on the data quality. We have checked that changing a reference band has very minor influence on the results. 

%%randomizes the flux measurements by their uncertainties (FR) and randomly chooses a subset of light-curve points (RSS) to build the peak/centroid lag distribution, whose median value and 1$\sigma$ range serve as the peak/centroid lag and error, respectively. We carried out 10000 FR/RSS iterations for each measurement.  

\subsubsection{Detrending for TDEs with RB bump}\label{sec:detrend}
The RB bump may be governed by different mechanisms from the power-law decay in TDE light curves. For example, according to the stream collision and delayed disk formation scenario, the observed UV/optical light curve could be a superposition of emissions from the shock and accretion disk, each governed by distinct mechanisms. To accurately measure the inter-band lags, it is necessary to perform the detrending \citep{Welsh99} to separate two components and avoid mutual interference in lag measurements, particularly when their flux contributions are comparable. It is not applicable to X-ray light curves, as these are generally believed to originate primarily from the accretion disk.

In this work, we define the RB bump as a feature showing significant flux excess relative to the decay trend during the fading stage. Seven TDEs are identified in Table \ref{tab:RB_infor} exhibiting a significant RB bump from previous literature. 

%rather than at the end of the decay, which may be mimicked by photometric residuals after subtracting the host component. 

%late-time AGN variability (e.g., AT 2018fyk in stage B in Figure \ref{fig:fyk}) or

We employed either a power-law or an exponential detrending in $t-\nu L_{\rm \nu}$ space for light curves in each band following previous studies \citep{vanVelzen21b}. Specifically, we first attempt to fit the light curve with a power-law function:
\begin{equation}
L_{\rm \nu} = L_{\rm \nu ,\ peak}\ \left( \frac{t - t_{\rm peak} +t_{\rm 0}}{t_{\rm 0}} \right)^{p},  \ \ t_{\rm turnover} > t >  t_{\rm peak}
\end{equation}
where $t$ is the rest frame time, $t_{\rm peak}$ is the light peak time, $t_{\rm turnover}$ is the time of the first turnover point of the concave light curve, and $t_{\rm 0}$ is the power-law normalization. $L_{\rm \nu ,\ peak}$ is the peak luminosity at frequency $\nu$. We simply focus on the decay regime ($t > t_{\rm peak}$) since our aim is to remove the potential contribution from other mechanism, e.g., shock. The power-law function is a widely used model for describing the decay phase of TDE and generally provides good fits for most cases. However, a few ZTF TDEs still exhibit obvious deviations from the power-law modeling \citep{Yao23}. Then, an exponential function is also used to fit all the light curves for comparison:
\begin{equation}
    L_{\rm \nu} = L_{\rm \nu ,\ peak}\ e^{-\frac{t - t_{\rm peak}}{\tau_{\rm decay}}}, \ \ t_{\rm turnover} > t >  t_{\rm peak}
\end{equation}
where $\tau_{\rm decay}$ is the decay time. 

We adopt a maximum likelihood approach to estimate the posterior distributions of our model parameters given the priors listed in Table \ref{tab:RB_prior}. To efficiently draw samples from the posterior probability distributions of the model parameters, we use {\tt emcee} \citep{Foreman-Mackey13}, a python implementation of the affine invariant ensemble sampler for Markov Chain Monte Carlo (MCMC). Similar to previous works \citep{Yao23,vanVelzen21b}, uncertainties of the light curves are amplified by a factor $f$ according to its intrinsic scatter in each target to complement the potentially underestimated photometric error. We use 100 walkers and 5000 steps, discarding the first 1500 steps to ensure convergence. Then we adopt the model with the best $\chi^{2}_{\nu} $ as our fiducial result as listed in Table \ref{tab:RB_detrending}. Note that we also examined the lag measurements with different detrending methods, and confirmed that the lag results are generally comparable within uncertainties.

\begin{figure*}
\centering
\includegraphics[width=16cm]{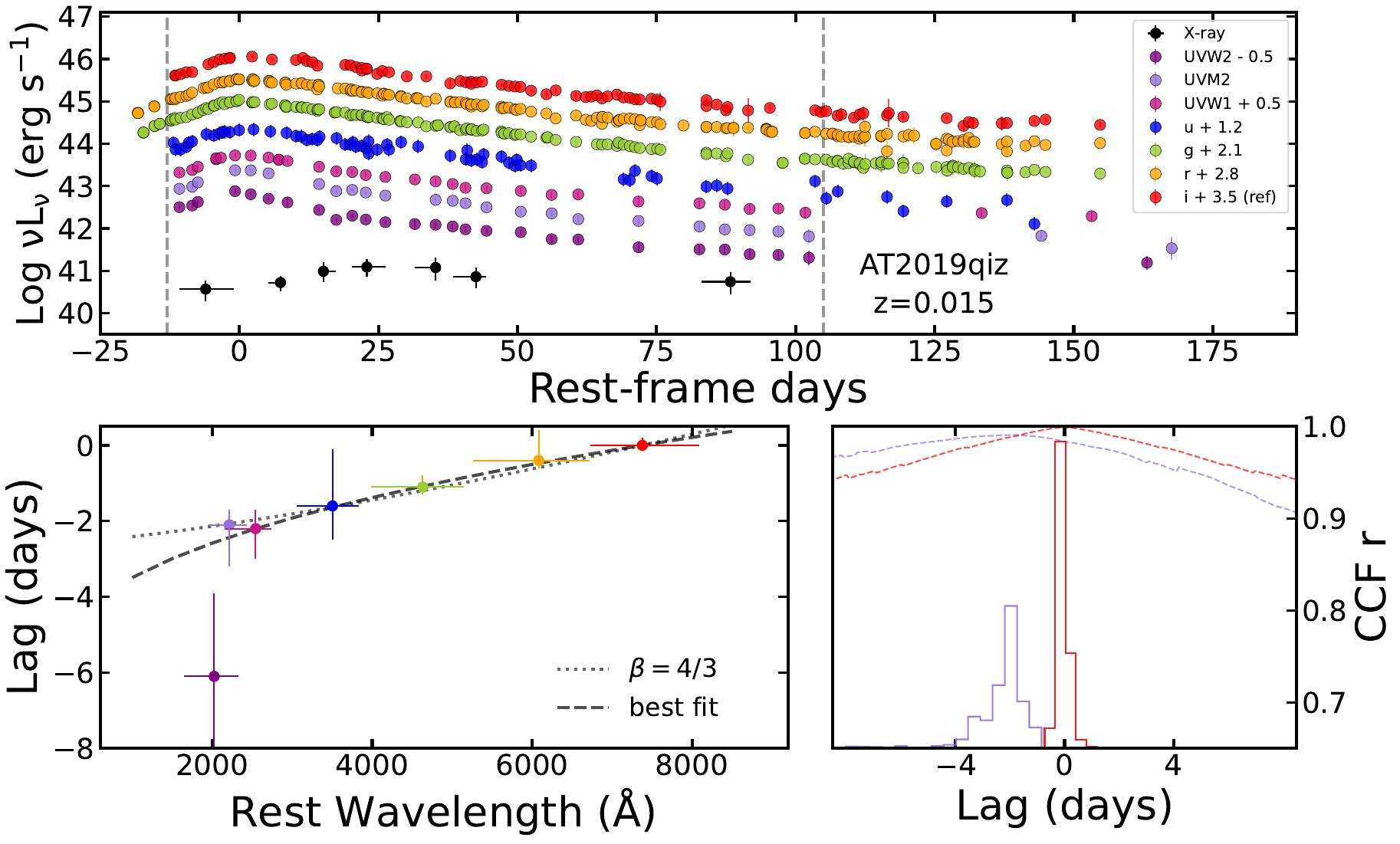}
\caption{AT 2019qiz multi-band light curves and ICCF lag measurements. Upper panel: combined light curves of UV/optical (host subtracted) and X-ray are obtained from Swift, ZTF, LCO, and Swope observations \citep{Nicholl20,Hung21}. The photometries used for lag measurements are bracketed by two grey dashed lines and all lags are measured relative to $i$ band. Bottom left panel: lag increases as a function of wavelength in the rest frame. The black dotted line represents the best power-law fit with a fixed slope of 4/3 ($\chi^2_{\rm \nu} = 0.43$) predicted by the standard thin disk model in AGN. The black dashed line represents the best fit with a free power-law slope ($\alpha = 0.4$, $\chi^2_{\rm \nu} = 0.39$). See more details in Appendix \ref{app:DC}. Bottom right panel: distributions of lag posteriors and cross-correlation curves for Swift UVM2-$i$ (light purple) and $i-i$ (red) bands in the rest frame (also see Figure \ref{fig:qiz_multi_lc} for other bands). }
\label{fig:2019qiz}
\end{figure*}

\begin{figure*}
\centering
\includegraphics[width=16cm]{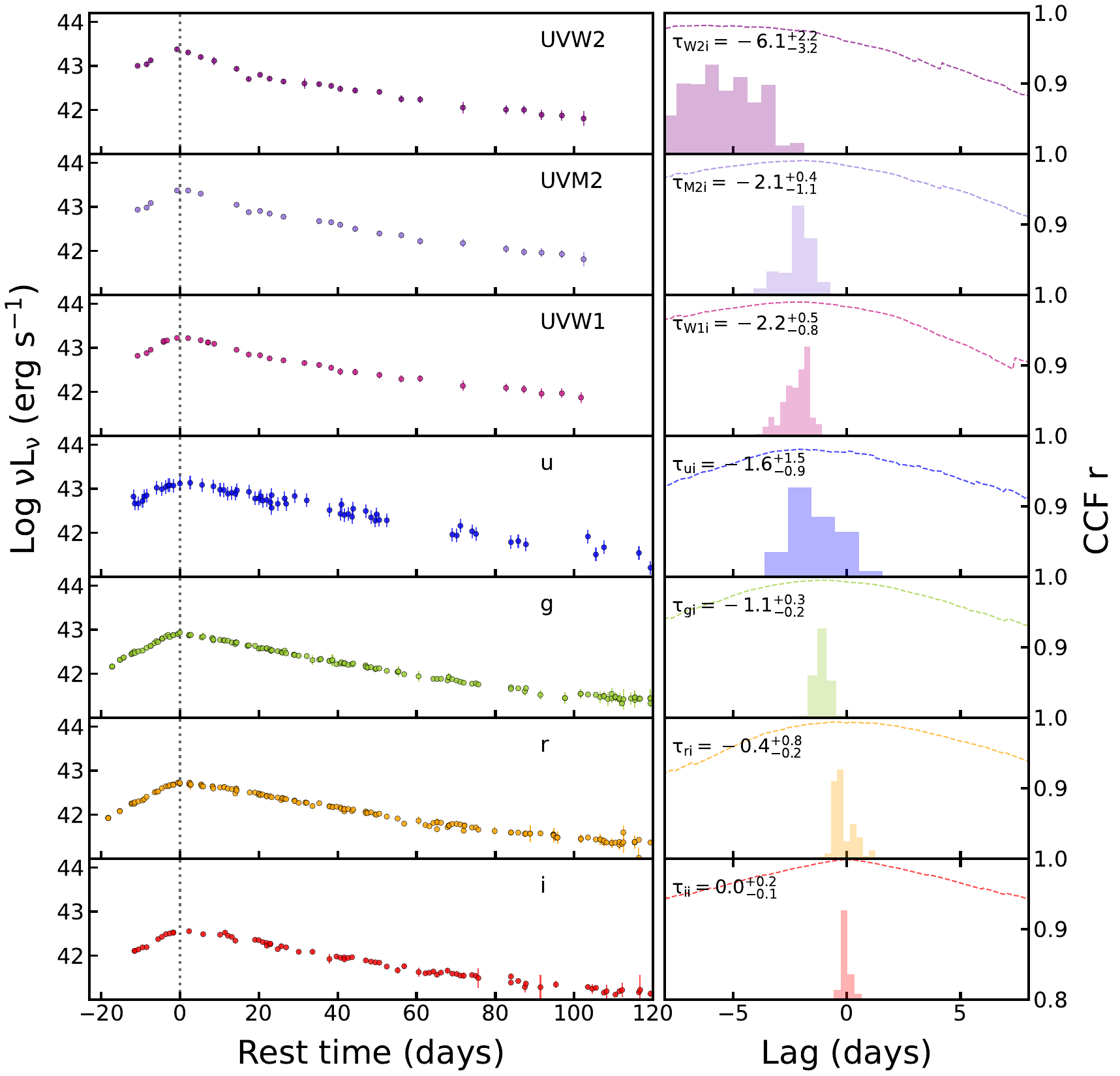}
\caption{{Multi-band light curves of AT2019qiz and ICCF lag measurements.} Left panel: multi-band light curves are observed by the Neil Gehrels Swift Observatory. Light curves are ordered by wavelength, from the shortest (top) to longest (bottom) wavelength. The $i$ band light curve is selected as the reference band. Its peak ($t=0$) is marked by a black dotted line. Right panel: ICCF lag posteriors based on Monte Carlo method with FR/RSS. The median values of these posteriors serve as the peak lags and their uncertainties are estimated from the 1$\sigma$ range of the distributions. The CCF curves (dashed lines) are shown in each panel. All their peak values ($r_{\rm max}$) are close to 1, indicating the strong correlations among these multi-band light curves. }
\label{fig:qiz_multi_lc}
\end{figure*}

\begin{figure*}
\centering
    \includegraphics[width=16cm]{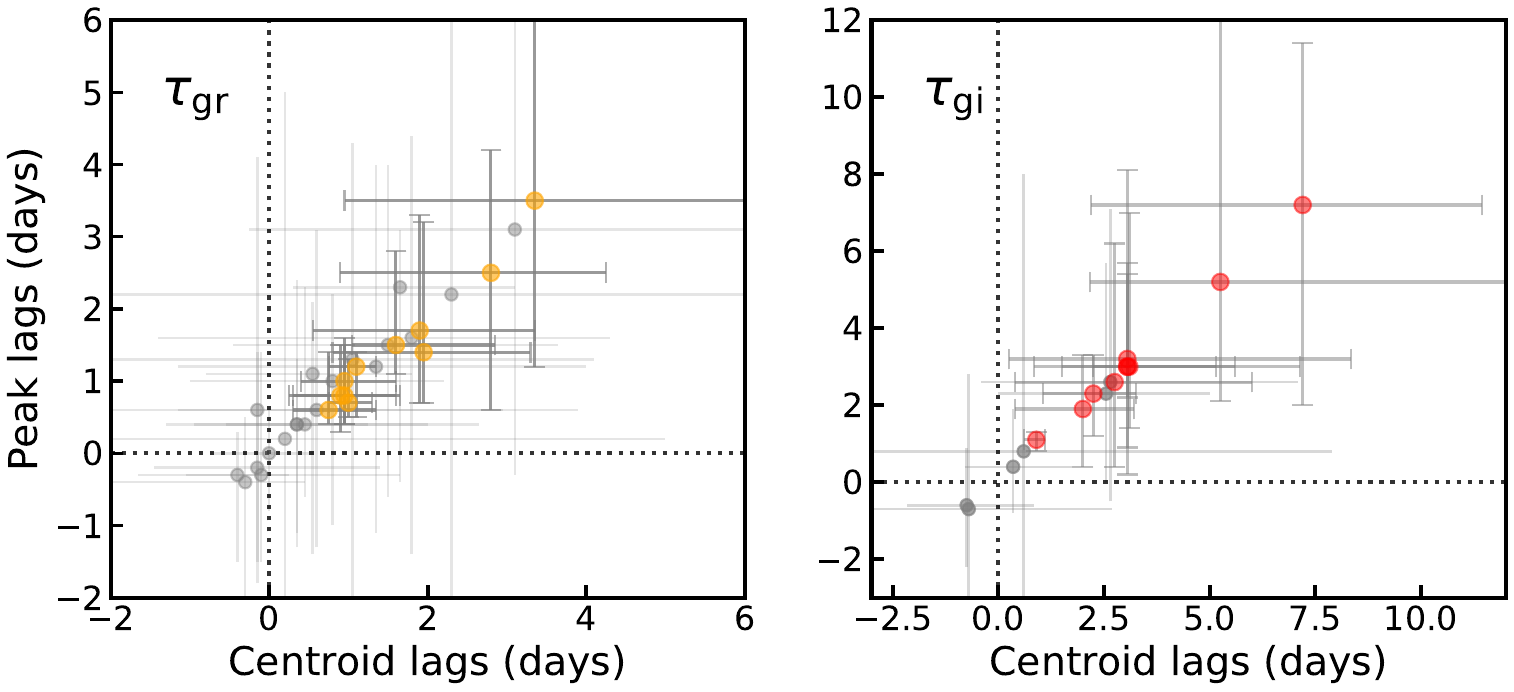}
    \caption{{Continuum lags for ZTF TDEs.}  Both centriod and peak lags measured with the ICCF method are exhibited, basically consistent with each other. $\tau_{\rm gi}$ are generally larger and more significant than the $\tau_{\rm gr}$ lags as expected. 11 $\tau_{\rm gr}$ (orange dots) and 9 $\tau_{\rm gi}$ (red dots) are larger than zero at $>$ 1$\sigma$ level. Grey dots represent the lags consistent with zero in errors. }
    \label{fig:stats}

\end{figure*}

\begin{figure*}
\includegraphics[width=17cm]{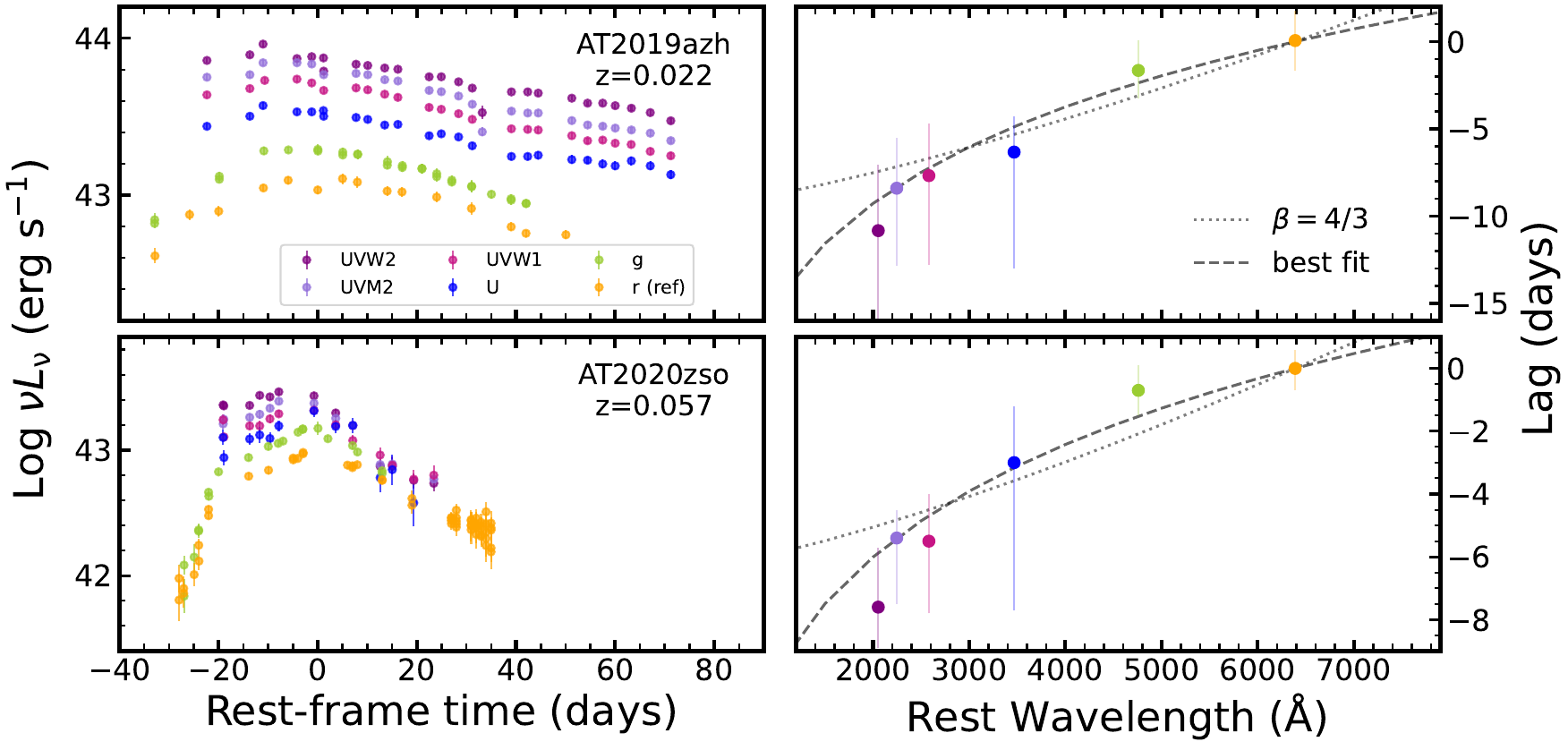}
\caption{{Multi-band TDE light curves and continuum lags in AT~2019azh and AT~2020zso.} Both objects show a few days of positive-direction continuum lags across the UV-optical band. The lags are measured relative to the $r$-band light curve. The lag model predicted by the standard thin disk theory (black dotted line) is clearly inconsistent with the observed lags, indicating the lag is not caused by the diffuse continuum. The dashed line is the best power-law fit. }
\label{fig:UV_lag}
\end{figure*}

\begin{figure}
%\centering
\includegraphics[width=8cm]{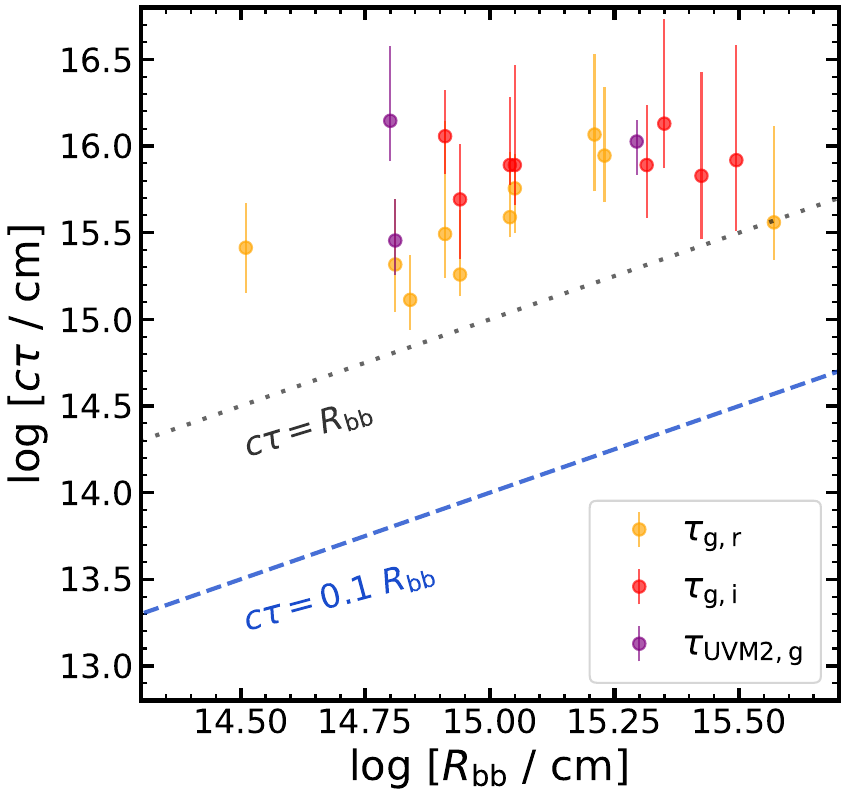}
\caption{Blackbody radius vs. continuum lag inferred size in TDEs. The total sample includes 16 sources, with 10 in $gr$ bands, 8 in $gi$ bands, and 3 in UV bands, each showing significance greater than 1$\sigma$. The dotted grey and dashed blue lines respectively show c$\tau$ = $R_{\rm bb}$, 0.1$R_{\rm bb}$ assuming a light-speed propagation.}
\label{fig:rbb}
\end{figure}

\section{Results and implications}\label{sec:results}
In this section, we systematically explore multi-band continuum variabilities across three distinct periods: the initial primary UV/optical peak, the RB bump, and the recurrent flare. We demonstrate that the results from reverberation mapping align more closely with scenarios involving stream collision and delayed disk formation, rather than envelope reprocessing. Consequently, we associate these three periods with specific physical processes: stream collision, disk formation, and rpTDE, respectively.

\subsection{Stream collision dominates the initial primary UV/optical peak}\label{sec:first_peak}

\subsubsection{Continuum reverberation in AT 2019qiz} 

Figure \ref{fig:2019qiz} shows the multi-band light curves and lag measurements of AT 2019qiz, where a single UV/optical flare dominates the overall variation in this TDE. The peak of the X-ray emission (0.3 $-$ 10 keV) is not detected until $\sim$ 30 days after the UV/optical peak, suggesting the X-ray emission is linked to the TDE rather than the original weak AGN accretion, which usually displays a positive-direction lag (high-energy band leads low-energy band) for fast-varying component \citep{McHardy14,Fausnaugh16,Cackett21}. In addition, a 30-day negative-direction lag is inconsistent with the inverse Compton scattering of NUV seed photons from the disk to X-ray in corona in AGN, which predicts a delayed X-ray of at most a few days \citep{Arevalo05,Pahari20}. 

Based on the high-cadence and multi-band monitoring, we detect robust positive-direction continuum lags, increasing as a function of wavelength across the UV-optical regime in Figure \ref{fig:2019qiz} \& \ref{fig:qiz_multi_lc}. The detected UV lag (UVM2) relative to optical ($i$) is $2.1^{+1.1}_{-0.4}$ days ($5.4^{+2.8}_{-1.0}$ $\times$ $10^{15}$ cm assuming light-speed transmission) with ICCF. This result indicates that the emitting distance between UV and optical has already exceeded the single-temperature-based blackbody radius ($R_{\rm bb}=6.5\times10^{14}$ cm for AT 2019qiz) by a factor of 8.3 at 4.8$\sigma$ in the envelope reprocessing scenario\footnote{We consider the light-speed transmission of the variability signal in TDE reprocessing to be a reasonable assumption, as is typically assumed in AGN reprocessing. Additionally, multi-temperature modeling of the spectral energy distribution (SED) may be more appropriate than a single-temperature blackbody model, which means the currently obtained blackbody temperature represents a lower limit, while $R_{\rm bb}$ is likely an upper limit.}. The tension will be further increased if the real separation between UV and optical emitting regions in the envelope is much smaller. Therefore, we suggest that the inter-band lags of days is difficult to be explained in the envelope reprocessing scenario \citep{Roth16,Dai18,Parkinson22,Thomsen22}, which usually predicts negligible lags in UV/optical (see \S \ref{sec:envelope}).

\subsubsection{Continuum reverberation of ZTF TDEs}
To confirm the positive-direction and day-level continuum lags between UV and optical, we also examined the ZTF sample consisting of 32 TDEs. As illustrated in Figure \ref{fig:stats}, most of the inter-band lags are positive yet the uncertainties are relatively large, subject to the ZTF's cadence and noise. In total, 11 $gr$ and 9 $gi$ lags significantly deviated from zero at $>1\sigma$ confidence level and two representative examples having $gri$ light curves are displayed in Figure \ref{fig:example}. More importantly, we also discovered two additional TDEs in Figure~\ref{fig:UV_lag} showing robust positive UV/optical lags increasing as a function of wavelength as AT 2019qiz, yielding a lag detection rate of 100\% between UV and optical bands given the ZTF cadence, albeit with a small sample size. The detected UV (UVW2) lags relative to optical ($r$) of AT 2019azh and AT 2020zso are $10.8^{+8.0}_{-3.8}$ days and $7.6^{+1.3}_{-1.9}$ days ($c\tau$ = $2.8^{+2.1}_{-1.0}\times 10^{16}$ cm and $2.0^{+0.3}_{-0.5}\times 10^{16}$ cm), exceeding their blackbody radii $R_{\rm bb}=5.6\times 10^{14}$ and 8.9$\times 10^{14}$ cm \citep{Hammerstein23} by factors of 50 at 2.7$\sigma$ and 22 at 3.8$\sigma$, respectively. This is indicative of a common occurrence of high-energy leading low-energy bands in the primary UV/optical peak.

Furthermore, we compared the blackbody radius ($R_{\rm bb}$) with the inferred size of the UV/optical emitting distance ($c\tau$ assuming light-speed transmission in reprocessing) for 16 sources showing significant lags above 1$\rm \sigma$. We remind that, according to the envelope reprocessing picture, the emitting separation across UV to optical should be much smaller than the whole envelope size, $c\tau \ll R_{\rm bb}$. Figure \ref{fig:rbb} reveals that the size inferred by $c\tau$ are generally larger than the entire envelope size ($R_{\rm bb}$). If we naively assume UV/optical emitting separation is around 0.1$R_{\rm bb}$, values of $c\tau$ in all targets are much larger than 0.1$R_{\rm bb}$. This finding further confirms that the current envelope-like structure is difficult to produce such a long lag between UV and optical emissions.

%We note that the envelope reprocessing model discussed here is still in its initial stages and may become more sophisticated over time. In addition, several factors might be oversimplified in current comparison. First, a multi-temperature model may serve as a better representation of the gas emission in UV/optical wavelengths than a single-temperature blackbody function \citep[e.g.][]{Liodakis2023,Dai18}. Second, the blackbody radius, assumed from a single-temperature model, is merely an effective approximation of the envelope size, rather than an accurate representation. Third, the actual structure may resemble an outflow or an asymmetric gas structure expanding radially, instead of simple spherical and concentric discrete shells \citep{Roth16}. 

%Finally, the TDE evolution within a reprocessing scenario involves various processes, such as adiabatic expansion, absorption/reemission, and Compton scattering, which may affect the assumptions based on light-speed transmission \citep{Thomsen22}.

\subsubsection{Stream collision}
In contrast, the delay of a few days between UV and optical emissions can be naturally explained by a discrete variant of the stream collision scenario. Given a penetration factor of $\beta$ $\sim$ 1 with relatively weak apsidal precession when the star passes the pericenter, it may produce a series of successive stream self-interactions \citep{Bonnerot17,Pasham17}. As depicted in episode II of Figure \ref{fig:cartoon}, a fast inner stream first collides with the infalling stream, generating UV emission. This is followed by a slower outer stream\footnote{Two streams may not necessarily be separated and different velocities are the key to produce UV and optical emissions. The width of the stream is strongly affected by the nozzle shock effect as it passes through the pericenter. We cannot completely rule out a relatively broad stream \citep[e.g., Figure~3b in ][]{Bonnerot21}, as it can also produce similar observed UV/optical lags, characterized by a fast-moving inner portion and a slower-moving outer portion.} that produces the optical emission, leading to a lag of a few days between the UV and optical emissions. 

%On the other hand, the approximate 30-day delay in the X-ray peak could be attributed to the formation of a new accretion disk, although no significant UV/optical rebrightening is observed in this object (see the disk formation in \S \ref{sec:second_peak}).

It is important to note that recent simulations have demonstrated that the material at the collision point is too optically thick to produce X-rays without adiabatic cooling \citep{Jiang16,Andalman22}, suggesting that the observed inter-band lags are unlikely to be caused by the X-ray reprocessing from the shock-powered outflows \citep{Lu20}. Additionally, we fitted the positive-direction lags (Figure \ref{fig:2019qiz} \& \ref{fig:UV_lag}) using both free and fixed slope power-law models. We propose that these lags are unlikely to originate from the diffuse continuum emission commonly observed in AGNs, considering the absence of high-energy photons and the lack of an evident $u/U$-band excess (see Appendix \ref{app:DC}). Furthermore, the continuum lag cannot be explained by changes in the size and temperature of the envelope over time, as this would predict dramatic UV-optical color variations, which is inconsistent with the relatively stable color observed in TDEs (see \S \ref{sec:envelope}).

\begin{figure*}
\centering
\includegraphics[width=16cm]{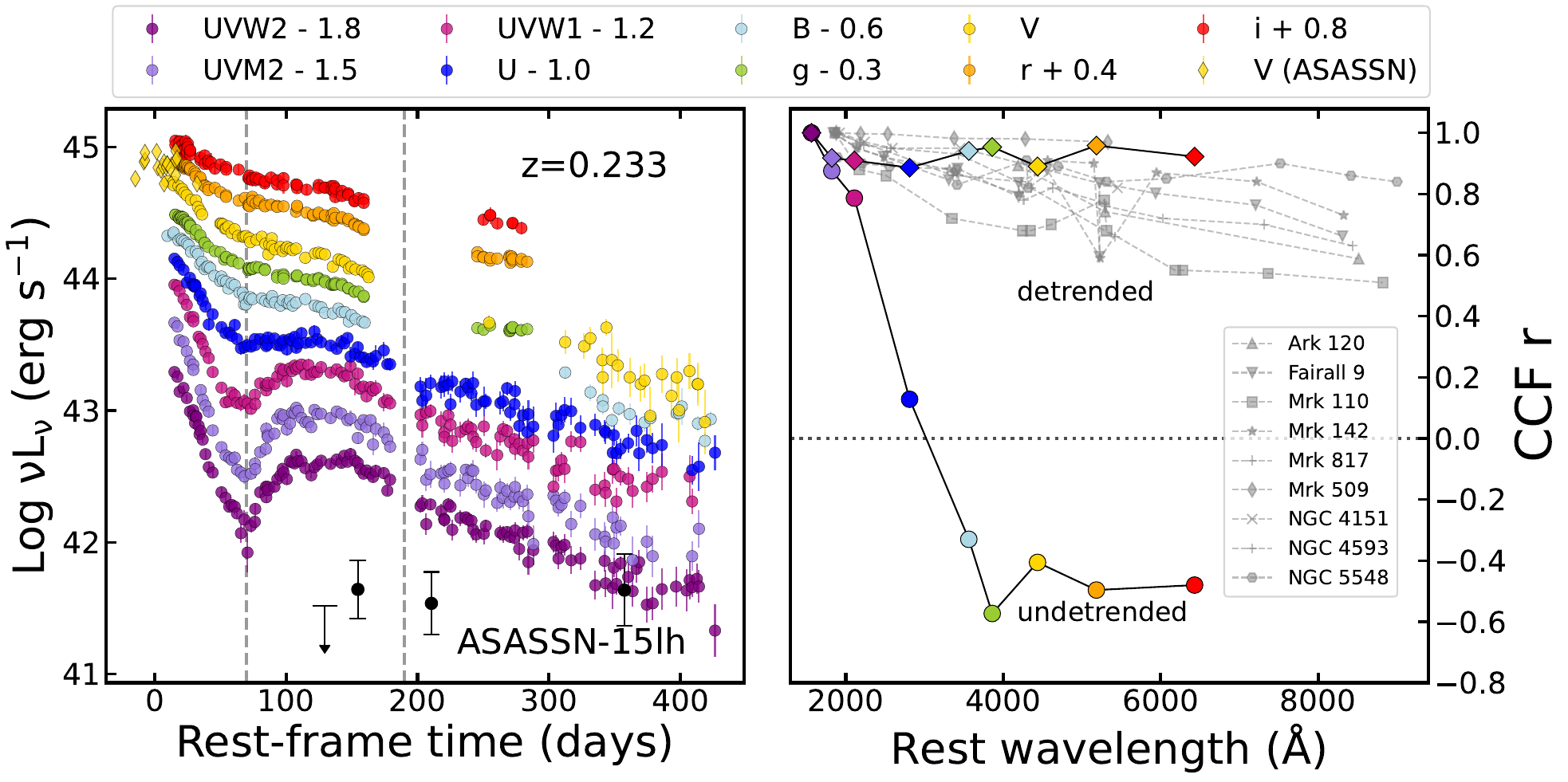}
\caption{Host-subtracted light curves of ASASSN-15lh and its inter-band cross correlations compared with local AGNs. Left panel: multi-band light curves obtained from LCO and Swift observations \citep{Leloudas16}. Black symbols (dot and arrow) represent the soft X-ray ($0.5-1$ keV) luminosity detected by Chandra, and the upper and lower error bars represent the X-ray luminosity adopting different models (power-law and blackbody functions). Right panel: inter-band cross-correlation coefficient curves (w.r.t. UVW2) of ASASSN-15lh and local AGNs (Akr 120, \citealt{Lobban20}, Fairall 9， \citealt{HS20}, Mrk 110, \citealt{Vincentelli21}, Mrk 142, \citealt{Edelson17}, Mrk 817, \citealt{Kara21}, Mrk 509, \citealt{Edelson19}, NGC 4151, \citealt{Edelson17}, NGC 4593, \citealt{Cackett18}, NGC 5548, \citealt{Fausnaugh16}) for comparison. The shock component is removed for the detrended version (see Figure \ref{fig:15lh_lag}). The ICCF analysis is based on the period of 70 to 190 days (grey dashed lines), where presents a significant anti-correlation ($r_{\rm max} \sim -0.6$) between the UV and optical.  }
\label{fig:15lh}
\end{figure*}

\subsection{Disk formation dominates the RB bump}\label{sec:second_peak}
\subsubsection{Two-component origin of the RB bump }
ASASSN-15lh, depicted in the left panel of Figure \ref{fig:15lh}, stands out as a target exhibiting prominent RB bumps with the best multi-band data quality to date. The short-term AGN-like variability also presents on the RB bump around 100 days after the light peak \citep{Margutti17}. The X-ray emission is first detected after the maximum of the RB peak and persists in $\sim$ 200 days, albeit with the sparse cadence. 

Cross-correlation analysis of the UV/optical light curves reveals that the UV and optical light curves around the RB bump ($\sim$ 100 days) are clearly anti-correlated with a maximum ICCF coefficient $r_{\rm max} \sim -0.6$ with respect to UVW2 in Figure \ref{fig:15lh} \citep[also see a similar case AT 2021ehb,][]{Yao22}. This is contrary to expectations of a strong positive correlation if the UV and optical emissions are driven by the same source (e.g., X-ray/EUV photons) in closely reprocessed layers with similar opacity, according to CRM results in AGNs \citep{Fausnaugh16,Lobban20,HS20,Cackett18,Vincentelli21,Kara21,Edelson17}.  Thus, a single reprocessing mechanism may not sufficiently explain the observed anti-correlation between UV and optical during the RB period.

%Instead the anti-correlation between X-ray and UV/optical that may be imitated by partial obscuration of the gas together with variations of ionizing luminosity. 
%the anti-correlated UV-optical light curves of the RB bump provide another piece of conclusive evidence against the pure reprocessing mechanism during the RB period.   

\begin{figure*}
\includegraphics[width=16cm]{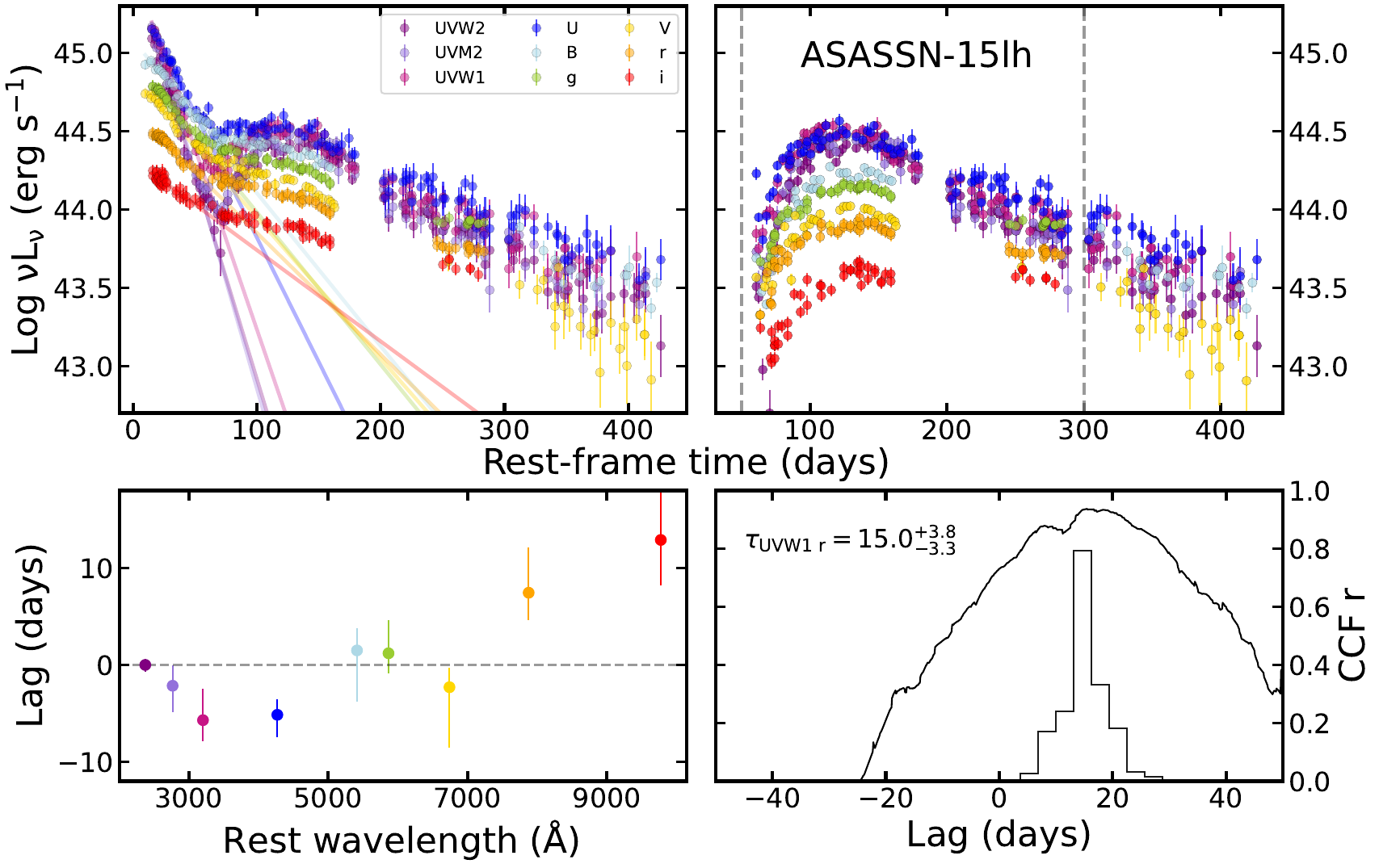}
\caption{{Light-curve detrending to separate the RB bump in ASASSN-15lh.} Upper panels: multi-band light curves before and after an exponential detrending. The region between two grey dashed lines is utilized for lag measurements. Bottom left panel: inter-band lags as a function of wavelength. Bottom right panel: lag posterior and CCF curve between UVW1 and $r$ band.}
\label{fig:15lh_lag}
\end{figure*}

\begin{figure*}
\includegraphics[width=16cm]{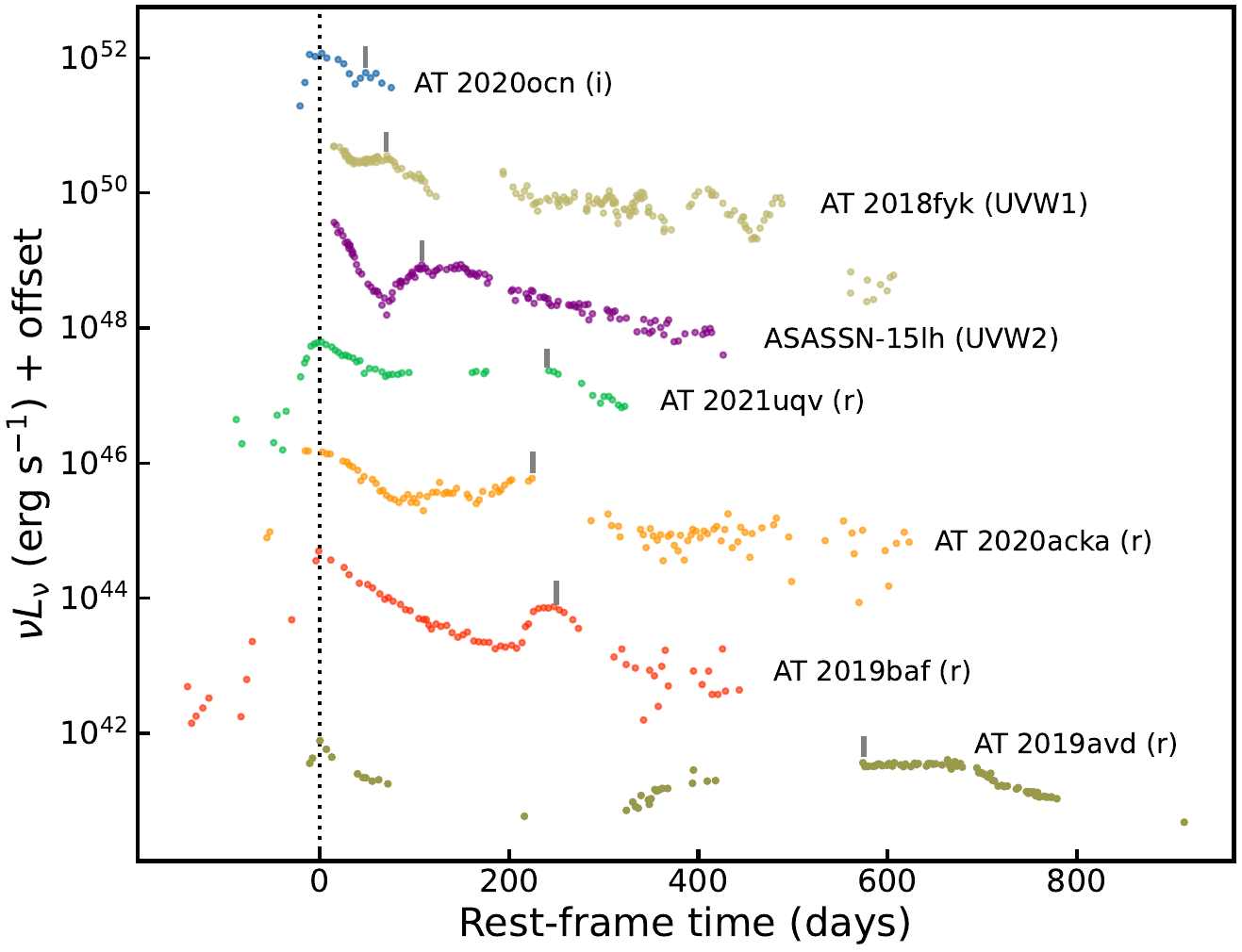}
\caption{{TDEs exhibiting prominent RB bumps reported in the literature.} Target names and corresponding bands are listed for each TDE. The first peak and the RB peak of the light curve are indicated by grey lines. Note that peak time of ASASSN-15lh was known by ASASSN V band in Figure \ref{fig:15lh}, and $t=0$ for AT 2018fyk represents the timing of its discovery, expected to be very close to the peak. AT 2019baf, AT 2020ocn, AT 2020acka, and AT 2021uqv are from a ZTF sample of 50 TDEs \citep{Hammerstein23,Yao23}, yielding an occurrence of 8\%.}
\label{fig:RB_sources}
\end{figure*}

Alternatively, if considering a two-component scenario, i.e., stream collision and delayed disk formation, the RB bump can be naturally explained by the disk formation that produces the X-ray/UV/optical emission \citep{Mummery20,Liu22}. Whether a certain TDE displays a UV/optical RB bump strongly depends on the relative contributions of the stream collision and disk emission. According to our understanding of disk formation (see \S \ref{sec:disk_formation}), the unique UV RB bump in ASASSN-15lh is likely caused by a significant amount of gas falling into the UV/EUV region whose emission peaks near UV regime and its contribution decreases with increasing wavelength. If we remove the underlying early-stage UV/optical contribution from the stream collision as shown in Figure \ref{fig:15lh_lag}, the multi-band variabilities of the accretion disk revert to positive correlations with coefficients generally decreasing with longer wavelength in the right panel of Figure \ref{fig:15lh}, similar to the results in AGN CRM: the UV and optical light curves are very tightly correlated and the inter-band lags increase as a function of wavelength as expected by X-ray reprocessing. 

%It is clear that the shock-contribution-subtracted coefficient curve ($r_{\rm max}$ as function of wavelength) follows a general decay trend similar to those in AGNs, supporting a disk origin of the RB bump in ASASSN-15lh. 

It is worth mentioning that the RB bump phenomenon in UV/optical light curve is not uncommon among detected TDEs. From a uniform, wide-field ZTF survey, 4 out of 50 TDEs exhibited a significant RB bump, resulting in an occurrence rate of 8\% (see Table \ref{tab:RB_detrending} and Figure \ref{fig:RB_sources}). This rate should be considered a lower limit, as many TDEs with ambiguous RB bumps were excluded from our simple evaluation.

\subsubsection{Lags of RB bumps} \label{sec:RB_lag}
Figure \ref{fig:RB_sources} presents seven TDEs exhibiting prominent RB bumps, which are valuable for the detailed investigation of the disk formation process. However, some objects, e.g., AT~2021uqv and AT~2020acka, lack multi-band light curves, thus are unavailable for reverberation analysis. In addition, no significant non-zero lags ($\tau_{\rm gr} = 0.4^{\rm +1.9}_{\rm -2.0}$) of the RB bump can be detected between $g$ and $r$ bands in AT 2019baf, albeit with high-cadence ZTF light curves. This probably indicates that the intrinsic lag is too small to be resolved by ZTF cadence. Therefore, in the following, we focus on four targets with significant non-zero lag detections.
%, namely ASASSN-15lh, AT 2019avd, AT 2020ocn, and AT 2018fyk. 
%We will discuss AT 2018fyk in \S \ref{sec:recurring}.

{\bf ASASSN-15lh:} As illustrated in Figure \ref{fig:15lh_lag}, the UV-optical lags associated with the RB bump generally exhibit a mild positive trend from 3000 to 10000\AA\ in the rest frame. However, this trend appears to reverse at wavelengths below 3000\AA, potentially due to relatively larger scatter in the UV light curves. By focusing specifically on the UVW1 and $r$ bands, which have a wide wavelength separation, we identified a significant lag of $15.0^{+3.8}_{-3.3}$ days (4.5$\sigma$). This suggests that the observed positive lag direction across the UV and optical wavelengths is likely genuine.

\begin{figure*} 
\includegraphics[width=16cm]{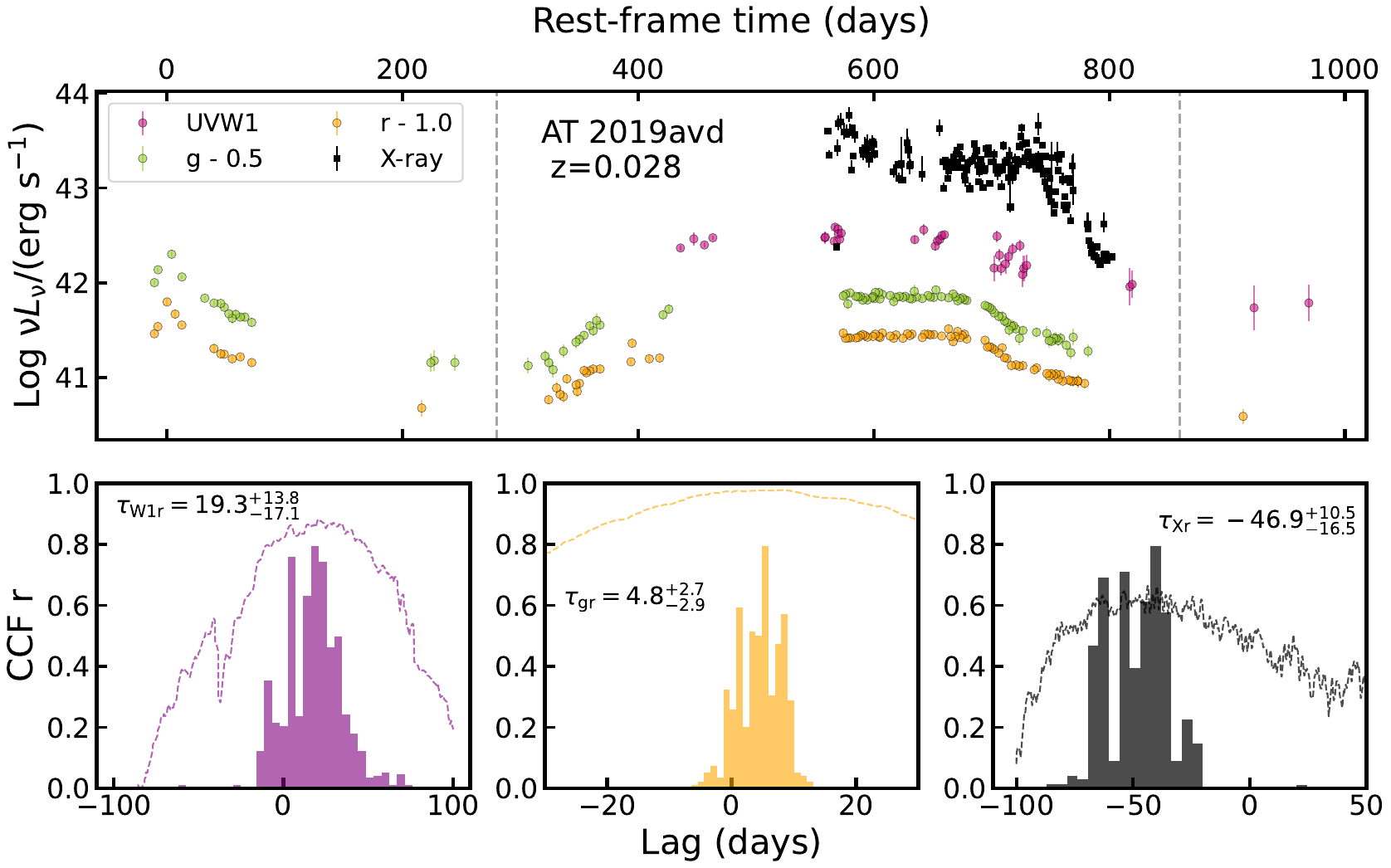}
\caption{Multi-band light curves and lag measurements of AT2019avd, a TDE with a broad RB bump. Upper panel: light curves are obtained from ZTF survey, Swift, and NICER (0.3-2 keV) telescopes \citep{Wang23}. Grey dashed lines bracket the RB region used for lag measurement after detrending. Bottom panels: inter-band lags are measured with the ICCF method. Positive-direction lags are detected in W1$r$ (1.1$\sigma$) and $gr$ (1.7$\sigma$) bands, albeit with larger  uncertainties. While a significant negative-direction lag (4.5$\sigma$) is detected between soft X-ray and $r$ band. CCF curves (dashed lines) and lag posteriors are plotted in each panel.}
\label{fig:avd}
\end{figure*}

{\bf AT 2019avd:} It was initially classified as a TDE candidate at $z = 0.028$ \citep{Malyali21} and subsequently confirmed as a genuine TDE \citep{Chen22,Wang23} upon revealing a significantly broad RB bump in Figure \ref{fig:avd} and other distinctive features observed during this bump, such as the \ion{He}{2} emission line, Bowen fluorescence lines, and high-ionization coronal lines. Most importantly, the significant negative-direction lag of $-46.9^{+10.5}_{-16.5}$ days (4.5$\sigma$) with $r_{\rm max} = 0.66$ between X-ray and optical bands further validates its TDE nature, distinguishing it from AGNs, which typically exhibit positive-direction lags. Note that the UV/optical turnover around 700 days is not related to the missing X-ray peak around 580 days, as indicated by additional Swift/XRT monitoring \citep{Wang23}. We propose that this may indicate an outside-in variability transition within the accretion flows on the nascent accretion disk (see \S \ref{sec:disk_formation}). In addition, mild positive-direction lags are also detected in UV/optical: $\tau_{\rm W1,r} = 19.3^{+13.8}_{-17.1}$ and $\tau_{\rm gr} = 4.8^{+2.9}_{-3.3}$ days. 

%Specifically, the gas initially fills the outer region of the accretion disk, which emits UV and optical light, before the accumulated gas spirals inward to fill the inner region that emits X-rays. For further details on disk formation, 

\begin{figure}
\hspace*{-0.5cm}
\includegraphics[width=9cm]{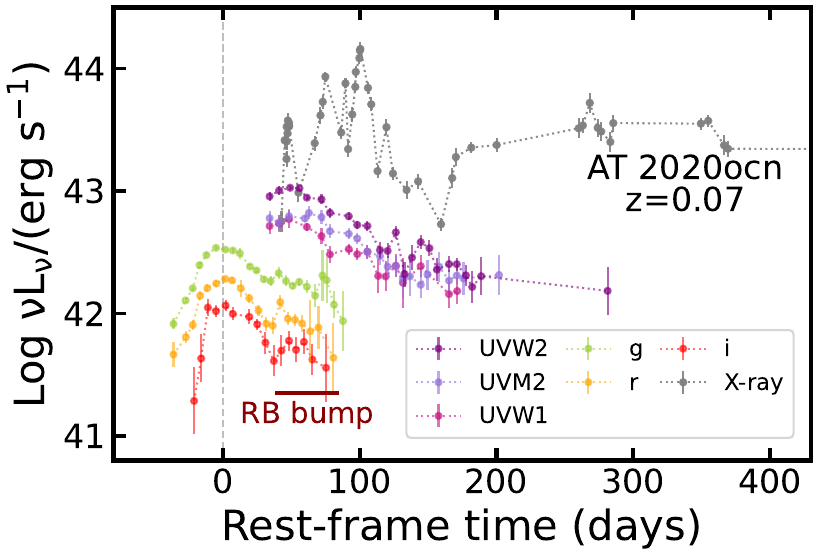}
\caption{{Multi-band light curves of AT 2020ocn with a relatively weak RB bump.} The host subtracted light curves are obtained from ZTF sample \citep{Hammerstein23}. The RB bump is observed $\sim$ 70 days after light maximum, although no UV data covers the first peak. The brown line indicates the RB bump and the grey dashed line indicates the first peak light. All the light curves are binned every 5 days for clarify. }
\label{fig:ocn}
\end{figure}

{\bf AT 2020ocn:} It is a spectroscopically confirmed TDE at $z = 0.07$ discovered by ZTF \citep{Hammerstein23}. As shown in Figure \ref{fig:ocn}, the optical monitoring reveals a relatively weak yet discernible RB bump occurring approximately 60 days after the primary peak. We consider this a robust RB bump since it corresponds to an observed UV peak, which is unlikely to be a delayed flare from the first peak. This is supported by the commonly observed positive-direction lags in shock-driven peaks, despite the lack of UV data covering the first peak. No significant lags among UV/optical bands are detected due to weak variability. Nevertheless, luminous X-ray emission is detected by Swift/XRT at $0.3-2$ keV, showing a clear delay of $42.4^{+3.4}_{-12.1}$ days relative to the RB bump in UV/optical. This again demonstrates the potential outside-in variability transition as AT 2020avd.

\begin{figure*}
\includegraphics[width=16cm]{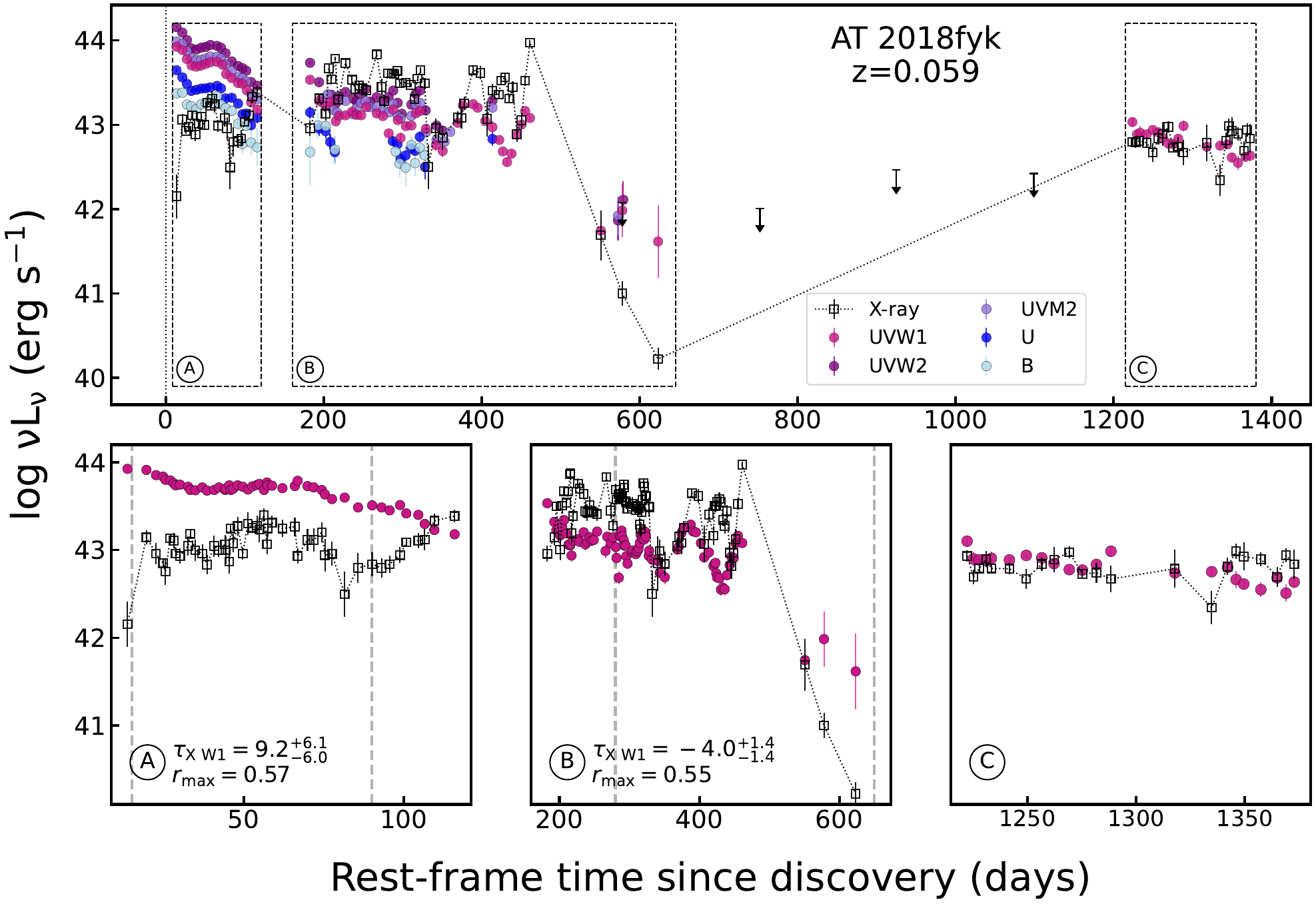}
\caption{{The long-term evolution of AT 2018fyk in UV/optical and X-ray.} Light curves are obtained from \citet{Wevers23}. Dots represent different host-subtracted UV/optical light curves and squares/arrows stand for the X-ray data including the upper limit. All the light curves are binned every 5 days for demonstration. Three periods in the upper panel are zoomed in showing in lower panels and in Figure \ref{fig:fyk_2stage}, and ICCF results are listed in each panel. Note that the UV and X-ray are clearly anti-correlated between $\sim$ 100 to 300 days, which are not considered for lag measurements.}
\label{fig:fyk}
\end{figure*}

{\bf AT 2018fyk:} It shows a distinct RB plateau in the UV/optical, even though the peak of the UV/optical flare is not captured in Figure \ref{fig:fyk}. The high-cadence, long-term monitoring enables us to investigate the X-ray/UV/optical correlations at different stages.

In stage A, the first detection of X-ray is in a low state but rises rapidly within 6 days, and then its variability pattern correlated with UV/optical for a period over $\sim$ 60 days. Later on, the UV/optical resumes to decline, whereas the X-ray gets further brightened showing an opposite variability trend between $100-300$ days after the peak, probably due to the state transition \citep{Wevers21}. Similar to ASASSN-15oi \citep{Holoien16}, the optical spectra in the RB plateau (at $\sim$ 45 days) display clear low-ionization $\rm Fe\ {\sc II}$ emission lines, indicative of optically thick and high-density gas, again providing compelling evidence of the formation of an optically thick accretion disk during the RB bump \citep{Wevers19a}. 

In stage B, both UV and X-ray present AGN-like variations with an almost flat trend rather than a TDE-like power-law decline, and eventually followed by a dramatic drop occurring in both UV and X-ray luminosity by several orders of magnitudes, indicating a sudden exhaustion of accretion flow in the disk after $\sim$ 500 days. The presence of low-ionization $\rm Fe\ {\sc II}$ emission lines ($\sim$ 45 days), along with a flat decay pattern accompanied with AGN-like variability ($\sim 200-400$ days) serve as corroborative evidence for the disk formation around the RB bump. 

\begin{figure*}
\centering
    \includegraphics[width=18cm]{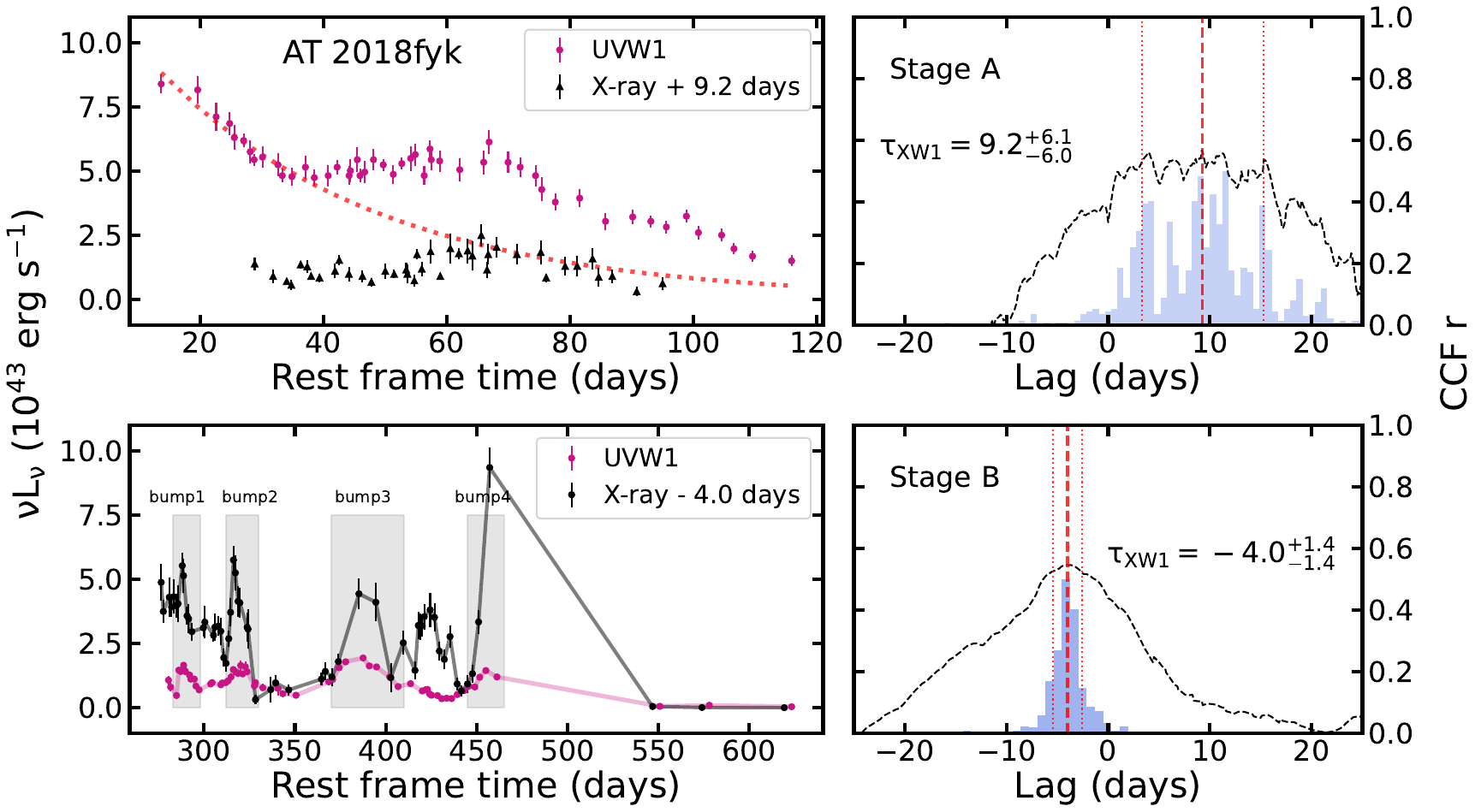}
    \caption{{Lag measurements of AT 2018fyk in stage A (upper panels) and stage B (lower panels).} Left panels: aligned UV and X-ray light curves. The X-ray light curve is shifted with the ICCF lag of $+9.2$ and $-4.0$ days, respectively. The selected light-curve segment may represent relatively pure mechanism and show the most significant CCF results. The shaded regions in the lower panel represent the corresponding bumps in two light curves. Right panel: lag posterior (blue histogram) and CCF curve (black dashed line). Red dashed and dotted lines denote the median value and 1$\sigma$ error, respectively.}
    \label{fig:fyk_2stage}
\end{figure*}

Remarkably, we detected opposite lag directions in these two stages (see Figure \ref{fig:fyk_2stage}): X-ray leads UV by $9.2^{+6.1}_{-6.0}$ days with an $r_{\rm max} = 0.57$ at stage A, whereas UV leads X-ray by $4.0^{+1.4}_{-1.4}$ days with an $r_{\rm max} = 0.55$ at stage B. These observations seemingly indicate different mechanisms of gas supplementation in different stages. According to the stream collision and delayed disk formation scenario, stage A is considered a very early formation phase of the disk where the variability is likely dominated by gas fallback. Meanwhile, stage B likely represents a relatively steady accretion phase, and its variability is likely to transition to being dominated by accretion flow on the disk. See more discussion of disk formation in \S \ref{sec:disk_formation}.

%However, we cannot rule out the possibility that the gas produced in the stream collision \citep{Lu20} and strong outflow caused by super-Eddington accretion are reprocessing the newly formed X-ray emission.

Therefore, we summarize several key findings based on these RB bump observations:
\begin{itemize}

\item The rise time and duration of the RB bump display a broad range, spanning from a few tens of days to several years. (Table \ref{tab:RB_infor}).

\item The observations suggest a bidirectional propagation of variability signals, with UV radiation slightly leading optical but significantly leading X-ray emissions (Table \ref{tab:RB_lag}).

\item The observed UV-X-ray lags are significantly longer than those typically seen in AGN reprocessing, and they also exhibit an opposite direction (Table \ref{tab:RB_lag}).

\item A potential switch in lag direction is noticed in AT~2018fyk, where a positive lag direction between X-ray and UV emissions is observed during the RB stage, transitioning to a negative direction during the subsequent accretion phase (Table \ref{tab:RB_lag}).
\end{itemize}

%\begin{figure*}
%\includegraphics[width=16cm]{fyk_stage1.pdf}
%\caption{{Lag measurements of AT 2018fyk in stage A.} Left panels: the exponential detrending (red dotted line) applied to the UVW1 light curve to remove the shock contribution. The X-ray light curve is aligned with the measured lags of 9.2 days with ICCF. The selected light-curve segment may represent relatively pure mechanism and show the most significant CCF results. Right panel: lag posterior (blue histogram) and CCF curve (black dashed line). Red dashed and dotted lines denote the median value and 1$\sigma$ error, respectively. }
%\label{fig:fyk_stage1}
%\end{figure*}

%\begin{figure*}
%\includegraphics[width=16cm]{fyk_stage2.pdf}
%\caption{{Lag measurements of AT 2018fyk in stage B.} Left panel: aligned UV and X-ray light curves. The X-ray light curve is shifted with the ICCF lag of $-4$ days. Shaded regions represent the corresponding bumps in two light curves. The selected light-curve segment may represent relatively pure mechanism and show the most significant CCF results. Right panel: lag posterior (blue histogram) and CCF curve (black dashed line). Red dashed and dotted lines denote the median value and 1$\sigma$ error, respectively.}
%\label{fig:fyk_stage2}
%\end{figure*}

%\begin{figure*}
%\includegraphics[width=16cm]{repeating_pTDE.pdf}
%\caption{An example of rpTDE candidate AT 2022dbl. The ZTF-r light curves (host subtracted) are binned with a 1-day window for clarity, and the original data are also shown with the transparent green dots. }
%\label{fig:pTDE}
%\end{figure*}

\begin{figure*}
\includegraphics[width=18cm]{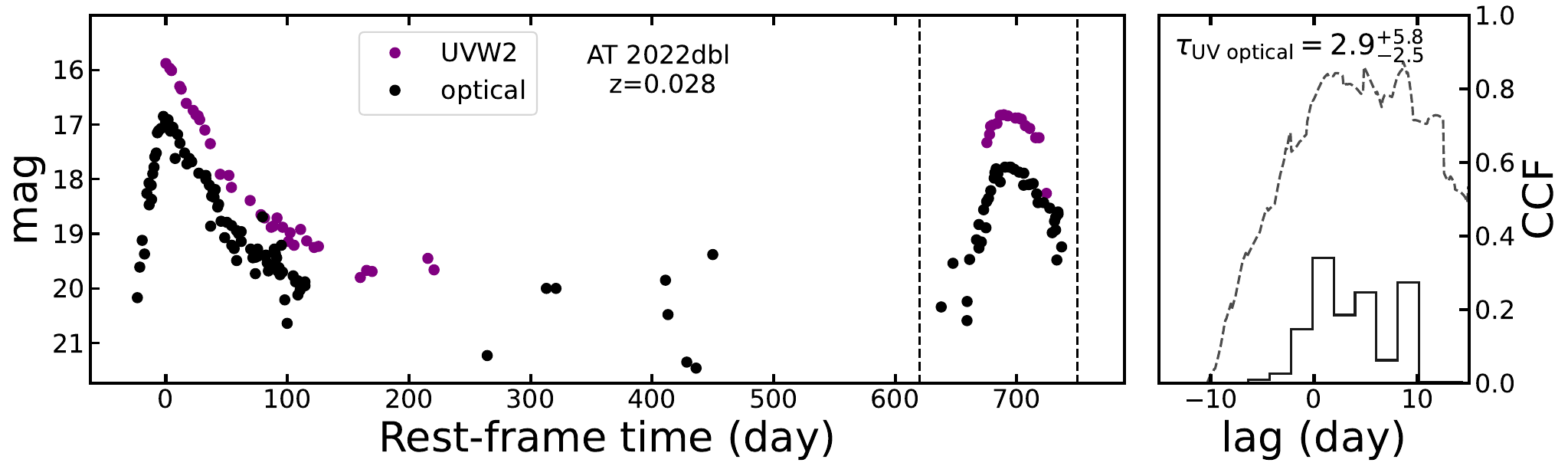}
\caption{An example of rpTDE candidate AT 2022dbl with lag measurements. The UV light curve corresponds to the Swift/UVW2 band, while the optical light curve is a combination of data from the LCO-$g$, ATLAS-$o$, and ATLAS-$c$ bands. The lag measurement of the second peak (region between two dashed lines) is shown in the right panel.}
\label{fig:pTDE}
\end{figure*}

\subsection{Recurrent flare}\label{sec:recurrent}

Returning to AT 2018fyk, the UV and X-ray emission at stage C remarkably recover to the level of stage B ($\sim$ 10\%$L_{\rm UV, peak}$). This recurrent flare plays a significant role in understanding the entire lifecycle of TDEs. Given the comparable X-ray and UV luminosities \citep{Wevers23}, such recurrent event indicates a new supplement of the accreting gas, probably resulting from a secondary stream collision in an rpTDE or sporadic accretion due to secondary outflow interactions \citep{Lu20}.

Likewise, other TDEs also demonstrated recurrent flares either in optical \citep[e.g., AT 2020vdq and AT 2022dbl,][]{Somalwar23,Lin24} or in X-ray \citep[eRASSt J045650.3$-$203750,][]{Liu23}. Occasionally, these flares display periodic behavior, with up to 30 cycles, as reported in ASASSN-14ko \citep{Payne21,Huang24}. Notably, AT 2022dbl, as shown in Figure \ref{fig:pTDE}, is considered the most robust example of an rpTDE, given the similar spectral features observed during the first and second flares, separated by an interval of approximately 740 days in the observed frame. In addition, its recurrent flare demonstrated a positive-direction lag of $2.9^{+5.8}_{-2.5}$ days between UV and combined optical light curves, indicative of a potential association with stream collision physics. If these recurrent flares indeed result from rpTDEs, the occurrence rate is non-trivial, and could be even higher than that of full TDEs \citep[e.g.,][]{Stone16}. Continued long-term monitoring of reported TDEs may lead to further discoveries of new rpTDEs.

\section{Theoretical interpretations}\label{sec:theory}
In this section, we first demonstrate that the inter-band lags of the primary UV/optical peak are difficult to explain within the envelope reprocessing scenario. Then, within the framework of the shock-disk scenario, we constrain the collision point based on the observed energy of the UV/optical emissions. Finally, we discuss the disk formation process based on our reverberation mapping results. 

\subsection{UV/optical lags in envelope reprocessing} \label{sec:envelope}
We explore two potential explanations for the inter-band lags observed in the initial primary UV/optical peak, within the context of the envelope reprocessing scenario.

\begin{figure}
\includegraphics[width=8cm]{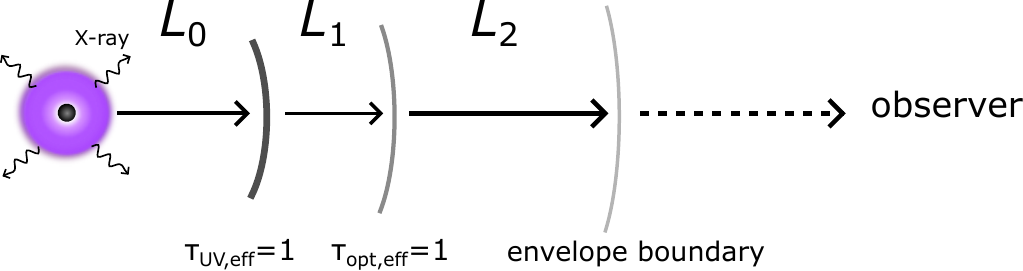}
\caption{{A simplified scheme of stratified TDE reprocessing model from \citep{Roth16}.} The accretion disk emitting soft X-ray is encapsulated by the quasi-spherical gas. Different layers represents the effective photospheres of UV, optical and electron scattering (envelope boundary). The effective photospheres at different bands refer to the boundaries beyond where most photons will electron scatter their way out of the envelope without being re-absorbed.  }
\label{fig:stratified_env}
\end{figure}

Previous work \citep[e.g.,][]{Roth16} proposed an extended, quasi-spherical, optically thick envelope, with a stratified structure where the effective photosphere for high-energy emissions (e.g., soft X-ray continuum emission) is located deeper within the envelope compared to the effective photosphere for low-energy emissions (e.g., UV/optical continuum emission). The effective photospheres ($\tau_{\rm eff} \sim \sqrt{\tau_{\rm es} \tau_{\rm abs}} =1$) at different wavelengths represent layers beyond which most photons can escape the envelope without being re-absorbed through electron scattering, as illustrated in Figure \ref{fig:stratified_env}. According to the reprocessing model, both the UV and optical emissions are driven by soft X-ray emission from the accretion disk and subsequently reprocessed by different layers of the envelope, with a negligible reprocessing timescale. To ensure the consistency with real observations, we assume that the envelope is sufficiently thick to absorb all soft X-ray photons and convert them into UV/optical emission. The time delay of UV/optical relative to X-ray variability strongly depends on the diffusion time (or electron scattering opacity, $\tau_{\rm es}$) through the envelope, namely the travel time between UV and optical photospheres to the outmost envelop boundary, i.e., $t_{\rm UV} = (L1\times \tau_{\rm L1,UV,es} + L2\times \tau_{\rm L2,UV,es} )/c$ and $t_{\rm opt} = (L1\times \tau_{\rm X-ray,L1,es} + L2\times \tau_{\rm opt,L2, es})/c$, respectively. $L1$ and $L2$ represent the separations between different shells in Figure \ref{fig:stratified_env}. While electron scattering opacity is determined by the scattering material, independent of the emission wavelength \citep{Paczynski83}, i.e., $\tau_{\rm L1,UV,es} = \tau_{\rm L1,X-ray,es}$ and $\tau_{\rm L2,UV,es} = \tau_{\rm L2,opt,es}$. Therefore, almost no theoretical time delay between the variability of UV and optical is expected through the path of $L1+L2$ in this simple scenario. Note that our calculations considered only very simple situations. The UV-optical delay necessitates further calculation if the reprocessing envelope is non-spherical or if the reprocessing is dominated by bound-free interactions in denser gas rather than by electron scattering.

In another scenario, it is possible that when the envelope size and effective photosphere temperature change with time, the peak of blackbody emission will also shift. Typically, when the luminosity and effective temperature decrease with time, the spectrum peak can move from the UV to the optical band, which can also cause a delay between UV and optical light curves. However, this mechanism will also cause significant evolution of spectrum slope with time. When the blackbody emission peaks at the UV band, we expect the total energy emitted in the UV band is larger than the total energy emitted in the optical band. When the peak moves to the optical band, the ratio should reverse. The real system can be more complicated if the effective temperature at the photosphere is not uniform. But the fact that the ratio between luminosity in UV and optical bands of TDE light curves shows very small variations with time suggests that this is also unlikely the explanation.

\subsection{Collision point} \label{sec:collision_point}
In the stream collision scenario, a critical parameter is the location of the collision point, which significantly influences the subsequent disk formation process. This collision point typically lies between the apocenter and pericenter and is primarily regulated by relativistic apsidal precession and the penetration depth of the stream \citep{Bonnerot21}. 

The expected photosphere temperature of the shocked gas from stream-stream collision is correlated with the distance $r_{\rm shock}$ to the BH where the collision happens. Assuming the stream is in a parabolic orbit, the velocity before the shock is $v_s=\left(2GM_{\rm BH}/r_{\rm shock}\right)^{1/2}$. Thickness of the stream $H_{\rm s}$ is typically determined by the balance between the pressure gradient inside the stream and the tidal force from the BH as $H_s\approx 2r_{\rm shock}\sqrt{r_{\rm shock}c_{\rm s}^2/(2r_{\rm g}c^2)} $ \citep{Jiang16}, where $c_{\rm s}$ is the sound speed of the stream and $r_{\rm g}$ is the gravitational radius of the BH. Notice that for a typical sound speed of $10$ km/s, thickness of the stream is smaller than $1\%r_{\rm g}$. Then typical density of the stream can be estimated as $\rho_{\rm s}=\dot{M}/(H_{\rm s}^2v_{\rm s})$, where $\dot{M}$ is the mass flux carried by the stream. The shock due to stream-stream collision converts the kinetic energy to the thermal energy. The temperature of the downstream gas $T_{\rm d}$ can be estimated by energy conservation $\rho_{\rm s} v_{\rm s}^2/2=3\rho_{\rm d} k_{\rm B}  T_{\rm d}/(2\mu m_{\rm p})+a_rT_{\rm d}^4$, where $k_{\rm B}$ and $a_r$ are  Boltzmann's constant and radiation constant respectively, $\rho_{\rm d}$ is the downstream gas density and we account for both gas internal energy and radiation energy density. The downstream gas will expand and cool, and the photosphere temperature will be smaller than $T_d$ by a factor of 20 or more depending on the $\dot{M}$ \citep{Jiang16}. Since wavelengths of UV and optical bands correspond to typical temperature of $2\times10^4$ and $10^4$ K, this requires the collision point to be smaller than $10^3r_{\rm g}$ for $10^6M_{\odot}$ BH and a few hundreds $r_{\rm g}$ for $10^8M_{\odot}$ BH. Otherwise the collision will not be able to produce hot enough gas for the prompt emission. Since the stream-stream collision point is likely the place where the disk starts to form, the estimated value of $r_{\rm shock}$ should set the outer edge of the disk.

\subsection{Disk formation}\label{sec:disk_formation}

\begin{figure*}
\includegraphics[width=16cm]{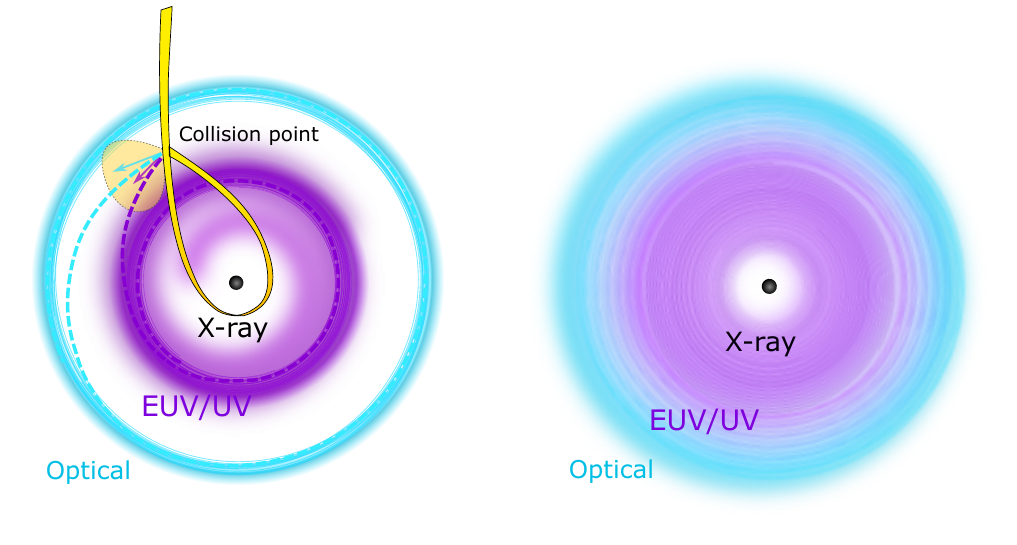}
\caption{A cartoon demonstrating the disk formation process: The left panel illustrates two representative inflowing and outflowing gas streams with different angular momenta circularizing to form EUV/UV and optical emitting annuli in the nascent disk. Since the gas angular momentum is continuously distributed after a shock collision, the right panel depicts a complete disk where the gaps are filled by spiraling or outflowing gas. The collision point is believed to be close to the UV/EUV emitting ring. }
\label{fig:disksize}
\end{figure*}

It is clear that the multi-band lags we observed here are governed by different physical mechanisms from those in AGNs \citep{Edelson15}. The lags are also significantly shorter compared with the viscous timescales predicted by the standard thin disk model if X-ray, UV and optical photons are produced by such a disk. In fact, even for normal AGNs, the thin disk model is far from successful to explain many observational properties of the accretion disks when the accretion rate is a few percent of the Eddington value \citep{KoratkarBlaes1999,DavisTchekhovskoy2020}. When the stream collides with itself in TDEs, the shock will not only convert the kinetic energy to the thermal energy, but also produce a broad distribution of angular momentum with respect to the BH for the downstream gas \citep{Jiang16,Lu20,Huang23}. The gas that is still bound to the BH will fall back and circularize at different radii according to the angular momentum they have. The gas that gains angular momentum during the collision will circularize at larger radii and emit radiation at a longer wavelength, while the gas that loses angular momentum during the process will circularize at smaller radii and emit radiation at a shorter wavelength, as shown in Figure \ref{fig:disksize}. Therefore, we propose a model of a small disk emitting EUV/UV to X-ray emission plus UV/optical rings outside.

The delayed X-ray is caused by the gradually spiral-in accretion flow from the EUV/UV emitting region (a few hundreds of $R_{\rm g}$) on an inflow timescale. During this processing, magnetic field may also play an important role in the formed accretion disk to reduce the inflow time compared with the estimated timescale based on the viscosity as used in the thin disk model. When the star is disrupted, stellar magnetic field will get stretched and be carried by the stream \citep{GuillochonMcCourt2017,Bonnerot+l2017}, which can be the seed magnetic field that gets amplified due to differential rotation when the disk is formed. If the seed magnetic field is in a favorable configuration (for example, with a poloidal component), strong toroidal magnetic field can be generated and elevate the accretion disk \citep{BegelmanSilk2017,DexterBegelman2019,BegelmanArmitage2023}. This type of accretion disk has a larger disk scale height compared with the standard thin disk model for the same accretion rate and therefore the inflow timescale is significantly reduced. Numerical simulations of isolated BH accretion disks have confirmed the existence of this type of accretion flow \citep{Jiang19+Corona,Mishra+2022,Huang+2023}. 

On the other hand, we observed robust UV-optical lags on RB bump (e.g., ASASSN-15lh). This requires extra UV/optical emitting rings outside the disk since the small disk truncated at UV/EUV emitting radius cannot directly produce the observed UV-optical lags and luminosities. This ring is produced by the gas gained angular momentum after the stream-stream collision or outflow gas. We emphasize that it represents a most effectively emitting ring, rather than totally isolated from the inner disk as the gas angular momentum is continuously distributed. The tight correlation between UV and optical could be due to the same gas origin from the collision point, e.g., similar mass distribution or intrinsic temperature fluctuation during the collision. The lag between UV and optical may correspond to the time it takes for the gas to circularize at different radii.  

For a Schwartzschild BH with a mass of $\sim 10^6 M_{\odot}$, the size of the accretion disk is relatively large in unit of $R_{\rm g}$. The fallback gas after circularization first generates the EUV emissions and then gradually spirals inward on a reduced inflow timescale, producing the soft X-rays (0.2 keV) with $T_{\rm peak}$ reaching up to $4\times 10^5$ K for a Schwarzschild BH ($T_{\rm peak} \approx 4 \times 10^5 \dot{m}^{1/4} (0.1/\eta)^{1/4} M_{\rm BH,6}^{-1/4}$, $\dot{m} = 1$, $\eta = 0.1$). Note that the gas fallback location does not necessarily correspond to the emitting peak of the SED. The observed negative-direction delay between X-ray and optical emission in AT2019avd and AT2020ocn indicate the location of fallback gas is close to the UV emitting annulus, rather than directly plunging into the X-ray emitting region. For the BH with a relatively larger mass, e.g., AT 2018fyk ($M_{\rm BH}=10^{7.7}M_{\odot}$), the disk size will be much smaller. The positive-direction lag between X-ray and UV may indicate the returning gases first fallback into the inner X-ray emitting region, then fill the EUV/UV zone or spread outwards on a diffuse timescale \citep{Shen14}. While the late-time negative-direction lag suggests a spiral-in accretion flow that occurs after the fallback gas becomes negligible. This pattern is similar to the slow variable component observed in AGNs, which is likely produced in the outer regions of the disk and propagates inward \citep{Pahari20}. For even larger BH masses ($\gtrsim 10^8$ M$_\odot$), the TDE could only occur in fast spinning BHs whose tidal radius is very close to the gravitational radius. This results in the formation of a very small accretion disk \citep{Strubbe09}. In addition, given a BH mass of $10^{8.5}$ $M_{\odot}$, the peak temperature ($T_{\rm peak}$) of the accretion disk is approximately $9.5 \times 10^4$ K, corresponding to the EUV band (at $\sim 300$ \AA). This explains why the UV RB bump in ASASSN-15lh is much more prominent than that in optical. The commonly observed positive-direction lags in UV/optical likely indicate the late formation of the optical emitting ring outside due to the longer circularization timescale. 

Therefore, we summarize some key theoretical findings in this section: 
\begin{itemize}

\item The UV/optical continuum lags of a few days in the first primary peak are difficulty to be explained in envelope reprocessing scenario.

\item The stream collision point should be sufficiently close to the BH. Otherwise, it will be unable to produce sufficient UV/optical luminosity to match the observations. 

\item Based on the reverberation results, we proposed a disk structure with a compact disk emitting X-ray in the center, surrounded by multiple gas rings (not physically isolated) emitting UV and optical emissions (see Figure \ref{fig:disksize}).  

\end{itemize}

\section{The unification model of TDE evolution}\label{sec:unificaiton}
\begin{figure*}
\centering
\includegraphics[width=16cm]{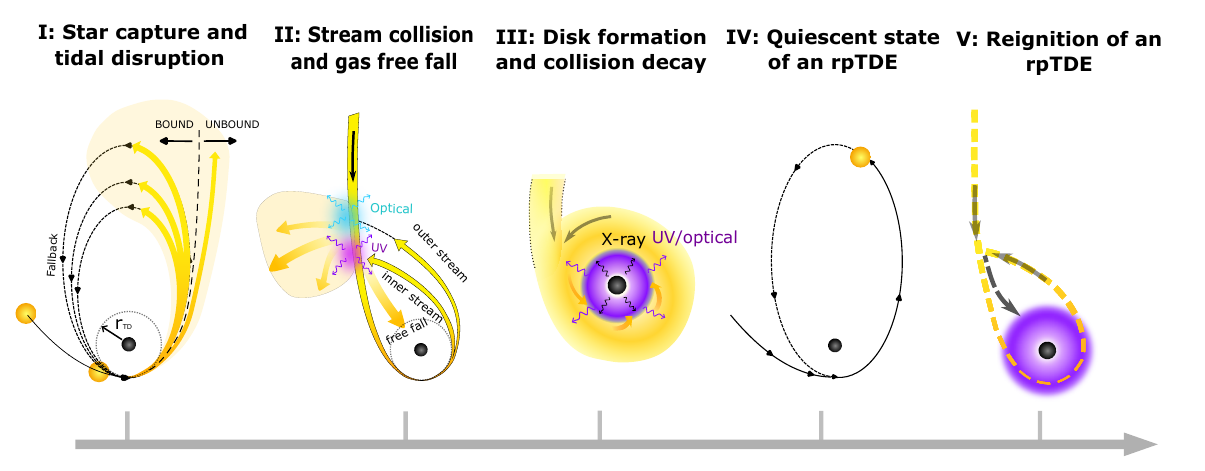}
\includegraphics[width=16cm]{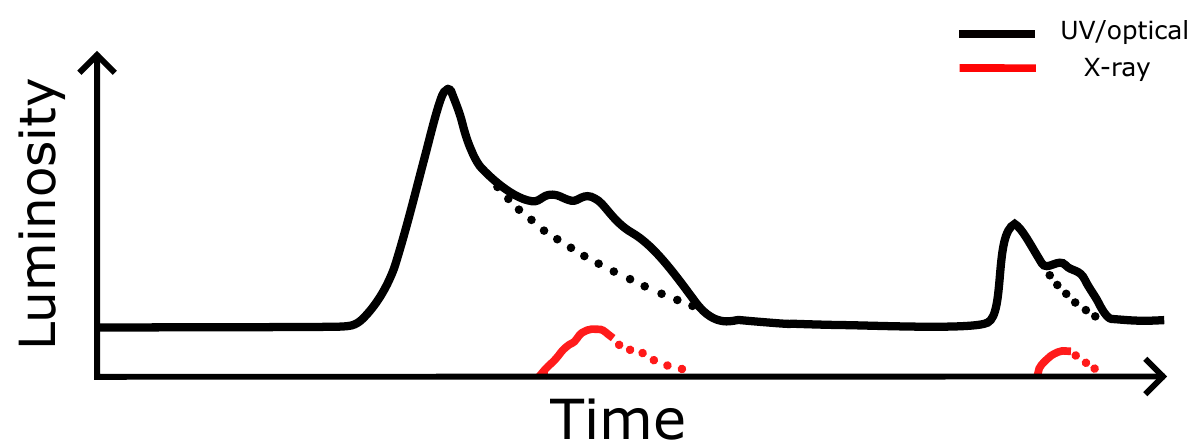}
\caption{A schematic of TDE evolution in stream collision and delayed disk formation scenario, and corresponding light curves. Upper panels: Episode I: a star is torn apart at the pericenter and around half of the stellar debris escapes while the rest material falls back to the BH. Episode II: because of the relativistic effects at pericenter, stream-stream collision occurs at a few hundred gravitational radius producing the primary UV/optical flare, an optically thick envelope, and strong outflows. The differential velocities of the discrete successive streams result in the observed continuum lags in UV/optical. Episode III: gas falls back to the BH and forms the new accretion disk emitting X-ray emission, accompanied by strong outflows. The stream collision gradually decays during this period. If the disk emission is significant, we may observe the X-ray emission and the RB bump in UV/optical light curves. Episode IV: the BH becomes quiescent again after all gases on the disk are accreted onto the BH. If the star is not fully disrupted and still bounded by the BH, it will fall back again into the tidal-disruption radius as an rpTDE. Episode V: the BH will be re-ignited as in episodes II and III, and a recurrent flare will be observed, accompanied by potential X-ray emission. Bottom panel: corresponding light curves in X-ray/UV/optical. The black dotted line indicates alternative evolution paths and the red dotted lines denote the uncertain late-time evolution of X-ray light curves.}
\label{fig:cartoon}
\end{figure*}

Based on the representative TDEs with lag information, we present a general evolutionary scheme for non-jetted TDEs in Figure \ref{fig:cartoon}, i.e., the stream collision and delayed disk formation scenario, assuming that all TDEs undergo a similar physical process.

A star is first disrupted by the tidal force of the massive BH at the pericenter in episode I. The properties of the outward stream debris are regulated by several relativistic effects \citep{Bonnerot21}, and the first primary UV/optical peak of the TDE flare is caused by the stream collision at a few hundred of gravitational radius. Due to the disparity in velocities in a discrete stream collision picture in episode II \citep{Bonnerot17,Pasham17}, the UV emission usually rises first followed by the optical emission with a lag of days. Meanwhile, an optically thick and anisotropic envelope will be produced after the stream collision, accompanied by strong outflows \citep{Jiang16,Huang23}. Moreover, if the outward stream interacts with materials in its path before colliding with the downstream, it may produce pre-peak bumps during its rising stage \citep[e.g., AT 2023lli,][]{Huang24}.

The gas energy dissipates/accumulates and angular momentum is redistributed during the stream collision in episode II. The gas may directly plunge into the massive BH on a freefall timescale or circularize until forming an accretion disk. During this stage, a RB bump in the UV/optical emission may be observed, comprising both shock emission (including shock-induced envelope/outflow emission) and radiation from the newly formed accretion disk. The presence and prominence of an obvious UV/optical RB bump strongly depend on the relative contributions of shock and disk emissions in episode III. The corona may be temporarily present, as suggested by the power-law component observed in the X-ray spectrum, which resembles the state transitions seen in stellar-mass black hole X-ray binaries. \citep{Yao22, Wevers21}. The shock emission usually dominates the early-stage emission of the TDEs and fades away with a power-law decay, thus highlighting the late-time disk emission with occasional AGN-like variability lasting for a viscous timescale. The late-time accretion is also confirmed by the long-standing FUV emission \citep{Vanvelzen19b} and X-ray detection \citep{Jonker20}. The delayed formation of the accretion disk naturally explains the delayed X-ray emission relative to the shock-induced UV/optical flare and their independent variabilities \citep[e.g.,][]{Yao22,Wevers23}. 

%On the other hand, due to the distinct physical origins of the shock-induced primary peak and the disk formation-induced RB bump, we anticipate that their power spectral densities (PSDs) should be distinguishable. In Appendix \ref{app:psd}, we found that the PSD slope of the RB bump is more tightly confined to $-2$ ($\propto \nu^{\alpha}$), similar to that observed in AGNs, whereas it is more divergent for the shock-induced peak. 

During the X-ray bright phase, X-ray reprocessing should contribute to the UV/optical luminosity if outer gas, produced by shock collisions or outflows, receives sufficient X-ray photons from the inner accretion disk or corona. We thus seek clues that indicate X-rays precede UV/optical emissions during the RB bump. However, significant observational evidence is still lacking, suggesting that X-ray reprocessing plays a limited role in the evolution of TDEs at least on current observations.

In addition, if a star is not fully disrupted, it will live another day, such as in a rpTDE in episode IV\&V, we may record a recurrent flare with a long-term monitoring \citep{Wevers23,Somalwar23,Lin24}. This intermittent ignition across optical-UV-X-ray also alleviates the long-standing missing energy problem \citep[e.g.,][]{Stone16,Lu18}.

\subsection{The power of unification: bridging diverse observations}
We then use our proposed unification model to elucidate the diverse observed light curves in UV/optical and X-ray. Generally, the TDE light curves can be categorized into three distinct groups:
\begin{itemize}
    \item Events that exhibit only optical/UV emission without detected X-ray emission \citep[e.g., most ZTF sample,][]{vanVelzen21b,Hammerstein23,Yao23}.
    
    \item Events that exhibit emissions in both optical/UV and X-rays \citep{Auchettl17,Guolo24}, yet with X-ray emissions displaying a wide lag range from a few days (e.g., ASASSN-14li, \citealt{Pasham17}, AT 2018fyk, \citealt{Wevers19a}, and AT 2019dsg, \citealt{Stein21}) to a few years (e.g., ASASSN-15oi, \citealt{Gezari17}, AT 2019azh, \citealt{Hinkle21}, and AT 2021ehb, \citealt{Yao22}).
    
    \item Events that primarily exhibit X-ray emissions with almost no optical/UV emission (e.g., cases discovered by ROSAT, \citealt{Bade96,Komossa99} and eROSITA, \citealt{Sazonov21}).
\end{itemize}
Typically, optically-selected TDEs, which have better multi-band light curve quality and immediate spectroscopic information, are considered relatively more robust than those identified by X-ray.

Naturally, if a TDE occurs and produces luminous UV/optical flare, it can be easily captured by modern wide-field, time-domain surveys. Whether it produces detectable X-ray emission strongly depends on the amount of gas falling onto the BH and contributing to disk formation. Based on current observations, the X-ray emission could have three appearances in UV/optical bright TDEs: no X-ray emission, slow X-ray emission (a few months or years delayed), and immediate X-ray emission (a few days delayed). 

After the stream collision, the gas loses angular momentum and falls back onto the BH. If the fallback gas is minimal or is directly swallowed by the BH without forming an accretion disk, it may lead to extremely faint X-ray emissions that are undetectable. On the other hand, if the fallback process is notably slow and inefficient at losing angular momentum, we expect a gradual increase in the X-ray emission, e.g., AT 2021ehb \citep{Yao22}. These X-ray light curves do not resemble the typical UV/optical light curve, which is characterized by a rapid rise followed by a gradual decay. Instead, they may display intricate potential variable features like multiple small peaks, plateaus, or sudden drops, reflecting the complexity of the disk formation process.

We suggest that the duration required to produce the disk X-ray emission depends on the location of the stream-stream collision point and the rate at which the gas loses angular momentum. If this timescale is as short as a few days to months, e.g., AT 2019dsg and ASASSN-14li, we will see prompt X-ray emission (immediate X-ray mode), yet still delayed to the UV/optical. Otherwise, if this timescale is slightly shorter than our long-term monitoring baseline ($\sim$ a few years), we may observe a long time delayed X-ray emission (slow X-ray mode). However, if it exceeds our monitoring duration or the X-ray flux falls below the detection limit, we might mistakenly identify it as a TDE lacking X-ray emission (no X-ray mode). Additionally, if a portion of the gas is drawn back onto the BH after passing the pericenter yet before the stream collision, we cannot rule out the possibility of observing X-ray emission preceding the UV/optical emission for the first primary peak.

Regarding X-ray bright-only TDEs, several factors could account for this phenomenon: the early UV/optical flare may have been missed by observations, the UV/optical flare might be relatively weak compared to the X-ray due to dust extinction, or the collision could be intrinsically weak, possibly due to misalignment of two streams \citep{Bonnerot21}. On the other hand, despite the high quality of UV/optical data, many TDE candidates remain indistinguishable and are instead classified as ambiguous nuclear transients \citep{Wiseman24}. This issue is likely even more severe when identifying X-ray TDEs, given the relatively poorer data quality.

It is worth noting that alongside optical bright and X-ray bright TDEs, there exists a population of IR bright TDEs, of which only a small fraction exhibit corresponding optical flares \citep{Jiang21a, Wang23, Masterson24}. Dust obscuration or intrinsic optical faintness may account for this discrepancy.   We suggest that this population will not significantly affect unification model of the TDE evolution as the IR emission mechanism is relatively clear.

%Our model investigates the regions surrounding black holes—within a few light-days—that emit X-ray, UV, and optical continuum emissions. The impact of a powerful TDE flare can propagate to more distant environments. For instance, the broad emission line region, which spans tens of light-days \citep{Nicholl20, Hinkle21}, the infrared-emitting dusty structure, ranging from a few to tens of light-months \citep{JiangN16, vanVelzen16a}, and the coronal line emitting regions, extending from several months up to a few light-years \citep{Komossa08, Wang12}.

%\section*{Disk formation dominates the rebrightening bump}
%ASASSN-15lh is a luminous transient ($z = 0.233$) discovered by ASASSN. It was initially classified as a superluminous supernova \citep{Dong16}, but subsequent evidence suggests it is a TDE originating from a fast-spinning Kerr BH with a mass of $\sim 10^{8.5} M_{\odot}$ \citep{Leloudas16,Margutti17,Gezari21}. A distinguishing hallmark is the significant rebrightening (RB) bump observed in the UV light curves whereas the optical light curve generally exhibit a power-law decay with a very weak RB trend. 

\section{Conclusions and Prospects}\label{sec:conclusion}
We systematically applied the continuum reverberation technique to approximately 30 ZTF TDEs, as well as to a select few TDEs with the best data quality to date. Our investigation aimed to accomplish four key scientific goals: 1) elucidate the lag properties of TDEs across different evolutionary stages; 2) understand the origin of UV/optical emissions; 3) distinguish between the shock-disk and the reprocessing scenarios; 4) explore a unified evolutionary TDE model that encapsulates the diverse classifications of TDEs. Overall, the reverberation results are more consistent with the shock-disk scenario (Figure \ref{fig:cartoon}), rather than the reprocessing scenario. The lag information also indicates that early-stage UV/optical emissions in TDEs are likely dominated by stream-stream collisions, while the nascent accretion disk may gradually become the predominant source over time. The specific findings are as follows:

\begin{itemize}
\item The initial primary peak of TDEs exhibits positive-direction UV/optical continuum lags with a few days. This lag inferred size is significantly larger than the envelope size or single-temperature blackbody radius in the reprocessing scenario. Moreover, the observed anti-correlation between UV and optical emissions in ASASSN-15lh \citep{Leloudas16}, provides further evidence challenging the current envelope reprocessing model. Alternatively, the stream collision scenario can naturally explain these observed UV/optical lags, given the different velocities in the discrete outward streams.

\item In addition to the lag and correlation evidence, other observations are also consistent with the shock-disk scenario. These include the frequently observed RB bump, accompanied by short-term, AGN-like variability \citep[ASASSN-15lh,][]{Margutti17}, and the low-ionization \ion{Fe}{2} emission lines associated with optically thick, high-density disk \citep[AT 2018fyk,][]{Wevers19a}. Further supporting features include the formation of the late-time corona, prolonged UV/X-ray emissions \citep{Jonker20,Vanvelzen19b}, and a high degree of polarization beyond this work \citep{Liodakis2023}.

\item The disk formation process may produce an RB bump in the UV/optical light curves during its decay phase. The presence of an RB bump in the UV/optical spectrum strongly depends on the flux ratio between the stream collision and the disk formation contributions. The lag between UV/optical and X-ray emissions during this stage is complex. We have identified significant cases, e.g., AT 2019avd (Figure \ref{fig:avd}), that exhibit negative-direction lags, with UV/optical emissions leading X-ray emissions by several tens of days. This probably indicates that the disk initially forms in the outer region (generating UV/optical emissions), and then the gas spirals inward to produce X-ray emissions, i.e., an outside-in formation of the accretion disk.

\item Generally, X-ray emission in TDEs typically lags behind the primary UV/optical peak ranging from days to years. It could also significantly lag behind the UV/optical RB bump if any up to a few tens of days. This could serve as a definitive characteristic of TDE flares, setting them apart from AGN transients, where X-ray generally precedes UV/optical continuum emission by a few days.

\item The proposed unified shock-disk scheme effectively accommodates the diverse optical and X-ray TDE populations, taking into account factors such as the efficiency of disk formation, the limitations of optical/X-ray observations, and intrinsic physics. 
\end{itemize}

It is worth mentioning that polarization observations can be a useful tool to distinguish between stream collision and reprocessing scenarios due to their distinct observational features \citep{Liodakis2023}. In the reprocessing scenario, the polarization degree is expected to be low ($\lesssim$ 10\%), as it arises from scattering in symmetric outflows. Conversely, polarization caused by synchrotron emission in stream collision could be several times higher, and the position angle may change with the shock. Therefore, future systematic polarization observations of TDEs will provide additional insights into the overall dynamics of TDE evolution.

On the other hand, the process of disk formation remains one of the most enigmatic aspects: whether it occurs from the outside-in, inside-out, or in a completely unordered manner. In the near future, as we observe an increasing number of TDEs exhibiting RB bumps and gather high-cadence data from instruments such as the deep fields in the Vera C. Rubin Observatory Legacy Survey of Space and Time \citep[LSST,][]{Zeljko19} and the Wide Field Survey Telescope \citep[WFST,][]{WangTG23}, complemented by observations from the Ultraviolet Transient Astronomy Satellite \citep[ULTRASAT,][]{Ben-Ami22} and the Einstein Probe \citep[EP,][]{Yuan22}, we expect to gain more profound insights into the disk formation process.

\begin{acknowledgements}
We thank the anonymous referee for helpful comments and suggestions. HXG and SLL thank Fukun Liu, Kirk Krista, Weimin Gu, Guobin Mou, and Yaping Li for useful discussions. HXG acknowledges support from the National Key R\&D Program of China No.~2022YFF0503402, 2023YFA1607903, and the National Natural Science Foundation of China (NFSC, No. 12473018). SLL is supported by the NFSC (No. 12273089). DFB is supported by the NSFC (No. 12173065). The Center for Computational Astrophysics at the Flatiron Institute is supported by the Simons Foundation. TGW is supported by the NSFC (No. 11833007) and the China Manned Space Project (CMS-CSST-2021-A12). YNW is supported by the Strategic Priority Research Program of the Chinese Academy of Sciences (No. XDB0550200). MYS acknowledges support from the NSFC (No. 12322303) and the Natural Science Foundation of Fujian Province of China (No. 2022J06002). MFG is supported by the Shanghai Pilot Program for Basic Research-Chinese Academy of Science, Shanghai Branch (JCYJ-SHFY-2021-013), the National SKA Program of China (Grant No. 2022SKA0120102), the science research grants from the China Manned Space Project with No. CMSCSST-2021-A06, and the Original Innovation Program of the Chinese Academy of Sciences (E085021002).
%\facility{FTN, FTS, DFOT, Palomar/P200}
%\software{AstroPy \citep{Astropy2018}, Linmix \citep{Brandon2007}, GALFIT \citep{Peng2002}, PyQSOFit \citep{Guo2018}}
\end{acknowledgements}

\bibliography{ms.bib}{}

\begin{thebibliography}{}
\expandafter\ifx\csname natexlab\endcsname\relax\def\natexlab#1{#1}\fi
\providecommand{\url}[1]{\href{#1}{#1}}
\providecommand{\dodoi}[1]{doi:~\href{http://doi.org/#1}{\nolinkurl{#1}}}
\providecommand{\doeprint}[1]{\href{http://ascl.net/#1}{\nolinkurl{http://ascl.net/#1}}}
\providecommand{\doarXiv}[1]{\href{https://arxiv.org/abs/#1}{\nolinkurl{https://arxiv.org/abs/#1}}}

\bibitem[{{Abramowicz} {et~al.}(1988){Abramowicz}, {Czerny}, {Lasota}, \& {Szuszkiewicz}}]{Abramowicz88}
{Abramowicz}, M.~A., {Czerny}, B., {Lasota}, J.~P., \& {Szuszkiewicz}, E. 1988, \apj, 332, 646, \dodoi{10.1086/166683}

\bibitem[{{Andalman} {et~al.}(2022){Andalman}, {Liska}, {Tchekhovskoy}, {Coughlin}, \& {Stone}}]{Andalman22}
{Andalman}, Z.~L., {Liska}, M. T.~P., {Tchekhovskoy}, A., {Coughlin}, E.~R., \& {Stone}, N. 2022, \mnras, 510, 1627, \dodoi{10.1093/mnras/stab3444}

\bibitem[{{Ar{\'e}valo} {et~al.}(2005){Ar{\'e}valo}, {Papadakis}, {Kuhlbrodt}, \& {Brinkmann}}]{Arevalo05}
{Ar{\'e}valo}, P., {Papadakis}, I., {Kuhlbrodt}, B., \& {Brinkmann}, W. 2005, \aap, 430, 435, \dodoi{10.1051/0004-6361:20041801}

\bibitem[{{Auchettl} {et~al.}(2017){Auchettl}, {Guillochon}, \& {Ramirez-Ruiz}}]{Auchettl17}
{Auchettl}, K., {Guillochon}, J., \& {Ramirez-Ruiz}, E. 2017, \apj, 838, 149, \dodoi{10.3847/1538-4357/aa633b}

\bibitem[{{Bade} {et~al.}(1996){Bade}, {Komossa}, \& {Dahlem}}]{Bade96}
{Bade}, N., {Komossa}, S., \& {Dahlem}, M. 1996, \aap, 309, L35

\bibitem[{{Begelman} \& {Armitage}(2023)}]{BegelmanArmitage2023}
{Begelman}, M.~C., \& {Armitage}, P.~J. 2023, \mnras, 521, 5952, \dodoi{10.1093/mnras/stad914}

\bibitem[{{Begelman} \& {Silk}(2017)}]{BegelmanSilk2017}
{Begelman}, M.~C., \& {Silk}, J. 2017, \mnras, 464, 2311, \dodoi{10.1093/mnras/stw2533}

\bibitem[{{Bellm} {et~al.}(2019){Bellm}, {Kulkarni}, {Graham}, {Dekany}, {Smith}, {Riddle}, {Masci}, {Helou}, {Prince}, {Adams}, {Barbarino}, {Barlow}, {Bauer}, {Beck}, {Belicki}, {Biswas}, {Blagorodnova}, {Bodewits}, {Bolin}, {Brinnel}, {Brooke}, {Bue}, {Bulla}, {Burruss}, {Cenko}, {Chang}, {Connolly}, {Coughlin}, {Cromer}, {Cunningham}, {De}, {Delacroix}, {Desai}, {Duev}, {Eadie}, {Farnham}, {Feeney}, {Feindt}, {Flynn}, {Franckowiak}, {Frederick}, {Fremling}, {Gal-Yam}, {Gezari}, {Giomi}, {Goldstein}, {Golkhou}, {Goobar}, {Groom}, {Hacopians}, {Hale}, {Henning}, {Ho}, {Hover}, {Howell}, {Hung}, {Huppenkothen}, {Imel}, {Ip}, {Ivezi{\'c}}, {Jackson}, {Jones}, {Juric}, {Kasliwal}, {Kaspi}, {Kaye}, {Kelley}, {Kowalski}, {Kramer}, {Kupfer}, {Landry}, {Laher}, {Lee}, {Lin}, {Lin}, {Lunnan}, {Giomi}, {Mahabal}, {Mao}, {Miller}, {Monkewitz}, {Murphy}, {Ngeow}, {Nordin}, {Nugent}, {Ofek}, {Patterson}, {Penprase}, {Porter}, {Rauch}, {Rebbapragada}, {Reiley}, {Rigault}, {Rodriguez}, {van Roestel}, {Rusholme}, {van
  Santen}, {Schulze}, {Shupe}, {Singer}, {Soumagnac}, {Stein}, {Surace}, {Sollerman}, {Szkody}, {Taddia}, {Terek}, {Van Sistine}, {van Velzen}, {Vestrand}, {Walters}, {Ward}, {Ye}, {Yu}, {Yan}, \& {Zolkower}}]{Bellm19}
{Bellm}, E.~C., {Kulkarni}, S.~R., {Graham}, M.~J., {et~al.} 2019, \pasp, 131, 018002, \dodoi{10.1088/1538-3873/aaecbe}

\bibitem[{{Ben-Ami} {et~al.}(2022){Ben-Ami}, {Shvartzvald}, {Waxman}, {Netzer}, {Yaniv}, {Algranatti}, {Gal-Yam}, {Lapid}, {Ofek}, {Topaz}, {Arcavi}, {Asif}, {Azaria}, {Bahalul}, {Barschke}, {Bastian-Querner}, {Berge}, {Berlea}, {Buehler}, {Dittmar}, {Gelman}, {Giavitto}, {Guttman}, {Haces Crespo}, {Heilbrunn}, {Kachergincky}, {Kaipachery}, {Kowalski}, {Kulkarni}, {Kumar}, {K{\"u}sters}, {Liran}, {Miron-Salomon}, {Mor}, {Nir}, {Nitzan}, {Philipp}, {Porelli}, {Sagiv}, {Schliwinski}, {Sprecher}, {De Simone}, {Stern}, {Stone}, {Trakhtenbrot}, {Vasilev}, {Watson}, \& {Zappon}}]{Ben-Ami22}
{Ben-Ami}, S., {Shvartzvald}, Y., {Waxman}, E., {et~al.} 2022, in Society of Photo-Optical Instrumentation Engineers (SPIE) Conference Series, Vol. 12181, Space Telescopes and Instrumentation 2022: Ultraviolet to Gamma Ray, ed. J.-W.~A. {den Herder}, S.~{Nikzad}, \& K.~{Nakazawa}, 1218105, \dodoi{10.1117/12.2629850}

\bibitem[{{Bonnerot} {et~al.}(2017{\natexlab{a}}){Bonnerot}, {Price}, {Lodato}, \& {Rossi}}]{Bonnerot+l2017}
{Bonnerot}, C., {Price}, D.~J., {Lodato}, G., \& {Rossi}, E.~M. 2017{\natexlab{a}}, \mnras, 469, 4879, \dodoi{10.1093/mnras/stx1210}

\bibitem[{{Bonnerot} {et~al.}(2017{\natexlab{b}}){Bonnerot}, {Rossi}, \& {Lodato}}]{Bonnerot17}
{Bonnerot}, C., {Rossi}, E.~M., \& {Lodato}, G. 2017{\natexlab{b}}, \mnras, 464, 2816, \dodoi{10.1093/mnras/stw2547}

\bibitem[{{Bonnerot} \& {Stone}(2021)}]{Bonnerot21}
{Bonnerot}, C., \& {Stone}, N.~C. 2021, Scientific Studies of Reading, 217, 16, \dodoi{10.1007/s11214-020-00789-1}

\bibitem[{{Cackett} {et~al.}(2021){Cackett}, {Bentz}, \& {Kara}}]{Cackett21}
{Cackett}, E.~M., {Bentz}, M.~C., \& {Kara}, E. 2021, iScience, 24, 102557, \dodoi{10.1016/j.isci.2021.102557}

\bibitem[{{Cackett} {et~al.}(2018){Cackett}, {Chiang}, {McHardy}, {Edelson}, {Goad}, {Horne}, \& {Korista}}]{Cackett18}
{Cackett}, E.~M., {Chiang}, C.-Y., {McHardy}, I., {et~al.} 2018, \apj, 857, 53, \dodoi{10.3847/1538-4357/aab4f7}

\bibitem[{{Chen} {et~al.}(2022){Chen}, {Dou}, \& {Shen}}]{Chen22}
{Chen}, J.-H., {Dou}, L.-M., \& {Shen}, R.-F. 2022, \apj, 928, 63, \dodoi{10.3847/1538-4357/ac558d}

\bibitem[{{Dai} {et~al.}(2018){Dai}, {McKinney}, {Roth}, {Ramirez-Ruiz}, \& {Miller}}]{Dai18}
{Dai}, L., {McKinney}, J.~C., {Roth}, N., {Ramirez-Ruiz}, E., \& {Miller}, M.~C. 2018, \apjl, 859, L20, \dodoi{10.3847/2041-8213/aab429}

\bibitem[{{Davis} \& {Tchekhovskoy}(2020)}]{DavisTchekhovskoy2020}
{Davis}, S.~W., \& {Tchekhovskoy}, A. 2020, \araa, 58, 407, \dodoi{10.1146/annurev-astro-081817-051905}

\bibitem[{{Dexter} \& {Begelman}(2019)}]{DexterBegelman2019}
{Dexter}, J., \& {Begelman}, M.~C. 2019, \mnras, 483, L17, \dodoi{10.1093/mnrasl/sly213}

\bibitem[{{Dong} {et~al.}(2016){Dong}, {Shappee}, {Prieto}, {Jha}, {Stanek}, {Holoien}, {Kochanek}, {Thompson}, {Morrell}, {Thompson}, {Basu}, {Beacom}, {Bersier}, {Brimacombe}, {Brown}, {Bufano}, {Chen}, {Conseil}, {Danilet}, {Falco}, {Grupe}, {Kiyota}, {Masi}, {Nicholls}, {Olivares E.}, {Pignata}, {Pojmanski}, {Simonian}, {Szczygiel}, \& {Wo{\'z}niak}}]{Dong16}
{Dong}, S., {Shappee}, B.~J., {Prieto}, J.~L., {et~al.} 2016, Science, 351, 257, \dodoi{10.1126/science.aac9613}

\bibitem[{{Edelson} {et~al.}(2015){Edelson}, {Gelbord}, {Horne}, {McHardy}, {Peterson}, {Ar{\'e}valo}, {Breeveld}, {De Rosa}, {Evans}, {Goad}, {Kriss}, {Brandt}, {Gehrels}, {Grupe}, {Kennea}, {Kochanek}, {Nousek}, {Papadakis}, {Siegel}, {Starkey}, {Uttley}, {Vaughan}, {Young}, {Barth}, {Bentz}, {Brewer}, {Crenshaw}, {Dalla Bont{\`a}}, {De Lorenzo-C{\'a}ceres}, {Denney}, {Dietrich}, {Ely}, {Fausnaugh}, {Grier}, {Hall}, {Kaastra}, {Kelly}, {Korista}, {Lira}, {Mathur}, {Netzer}, {Pancoast}, {Pei}, {Pogge}, {Schimoia}, {Treu}, {Vestergaard}, {Villforth}, {Yan}, \& {Zu}}]{Edelson15}
{Edelson}, R., {Gelbord}, J.~M., {Horne}, K., {et~al.} 2015, \apj, 806, 129, \dodoi{10.1088/0004-637X/806/1/129}

\bibitem[{{Edelson} {et~al.}(2017){Edelson}, {Gelbord}, {Cackett}, {Connolly}, {Done}, {Fausnaugh}, {Gardner}, {Gehrels}, {Goad}, {Horne}, {McHardy}, {Peterson}, {Vaughan}, {Vestergaard}, {Breeveld}, {Barth}, {Bentz}, {Bottorff}, {Brandt}, {Crawford}, {Dalla Bont{\`a}}, {Emmanoulopoulos}, {Evans}, {Figuera Jaimes}, {Filippenko}, {Ferland}, {Grupe}, {Joner}, {Kennea}, {Korista}, {Krimm}, {Kriss}, {Leonard}, {Mathur}, {Netzer}, {Nousek}, {Page}, {Romero-Colmenero}, {Siegel}, {Starkey}, {Treu}, {Vogler}, {Winkler}, \& {Zheng}}]{Edelson17}
{Edelson}, R., {Gelbord}, J., {Cackett}, E., {et~al.} 2017, \apj, 840, 41, \dodoi{10.3847/1538-4357/aa6890}

\bibitem[{{Edelson} {et~al.}(2019){Edelson}, {Gelbord}, {Cackett}, {Peterson}, {Horne}, {Barth}, {Starkey}, {Bentz}, {Brandt}, {Goad}, {Joner}, {Korista}, {Netzer}, {Page}, {Uttley}, {Vaughan}, {Breeveld}, {Cenko}, {Done}, {Evans}, {Fausnaugh}, {Ferland}, {Gonzalez-Buitrago}, {Gropp}, {Grupe}, {Kaastra}, {Kennea}, {Kriss}, {Mathur}, {Mehdipour}, {Mudd}, {Nousek}, {Schmidt}, {Vestergaard}, \& {Villforth}}]{Edelson19}
---. 2019, \apj, 870, 123, \dodoi{10.3847/1538-4357/aaf3b4}

\bibitem[{{Evans} \& {Kochanek}(1989)}]{Evans89}
{Evans}, C.~R., \& {Kochanek}, C.~S. 1989, \apjl, 346, L13, \dodoi{10.1086/185567}

\bibitem[{{Fausnaugh} {et~al.}(2016){Fausnaugh}, {Denney}, {Barth}, {Bentz}, {Bottorff}, {Carini}, {Croxall}, {De Rosa}, {Goad}, {Horne}, {Joner}, {Kaspi}, {Kim}, {Klimanov}, {Kochanek}, {Leonard}, {Netzer}, {Peterson}, {Schn{\"u}lle}, {Sergeev}, {Vestergaard}, {Zheng}, {Zu}, {Anderson}, {Ar{\'e}valo}, {Bazhaw}, {Borman}, {Boroson}, {Brandt}, {Breeveld}, {Brewer}, {Cackett}, {Crenshaw}, {Dalla Bont{\`a}}, {De Lorenzo-C{\'a}ceres}, {Dietrich}, {Edelson}, {Efimova}, {Ely}, {Evans}, {Filippenko}, {Flatland}, {Gehrels}, {Geier}, {Gelbord}, {Gonzalez}, {Gorjian}, {Grier}, {Grupe}, {Hall}, {Hicks}, {Horenstein}, {Hutchison}, {Im}, {Jensen}, {Jones}, {Kaastra}, {Kelly}, {Kennea}, {Kim}, {Korista}, {Kriss}, {Lee}, {Lira}, {MacInnis}, {Manne-Nicholas}, {Mathur}, {McHardy}, {Montouri}, {Musso}, {Nazarov}, {Norris}, {Nousek}, {Okhmat}, {Pancoast}, {Papadakis}, {Parks}, {Pei}, {Pogge}, {Pott}, {Rafter}, {Rix}, {Saylor}, {Schimoia}, {Siegel}, {Spencer}, {Starkey}, {Sung}, {Teems}, {Treu}, {Turner}, {Uttley}, {Villforth},
  {Weiss}, {Woo}, {Yan}, \& {Young}}]{Fausnaugh16}
{Fausnaugh}, M.~M., {Denney}, K.~D., {Barth}, A.~J., {et~al.} 2016, \apj, 821, 56, \dodoi{10.3847/0004-637X/821/1/56}

\bibitem[{{Foreman-Mackey} {et~al.}(2013){Foreman-Mackey}, {Hogg}, {Lang}, \& {Goodman}}]{Foreman-Mackey13}
{Foreman-Mackey}, D., {Hogg}, D.~W., {Lang}, D., \& {Goodman}, J. 2013, \pasp, 125, 306, \dodoi{10.1086/670067}

\bibitem[{{Gaskell} \& {Sparke}(1986)}]{Gaskell86}
{Gaskell}, C.~M., \& {Sparke}, L.~S. 1986, \apj, 305, 175, \dodoi{10.1086/164238}

\bibitem[{{Gezari}(2021)}]{Gezari21}
{Gezari}, S. 2021, \araa, 59, 21, \dodoi{10.1146/annurev-astro-111720-030029}

\bibitem[{{Gezari} {et~al.}(2017){Gezari}, {Cenko}, \& {Arcavi}}]{Gezari17}
{Gezari}, S., {Cenko}, S.~B., \& {Arcavi}, I. 2017, \apjl, 851, L47, \dodoi{10.3847/2041-8213/aaa0c2}

\bibitem[{{Godoy-Rivera} {et~al.}(2017){Godoy-Rivera}, {Stanek}, {Kochanek}, {Chen}, {Dong}, {Prieto}, {Shappee}, {Jha}, {Foley}, {Pan}, {Holoien}, {Thompson}, {Grupe}, \& {Beacom}}]{Godoy-Rivera17}
{Godoy-Rivera}, D., {Stanek}, K.~Z., {Kochanek}, C.~S., {et~al.} 2017, \mnras, 466, 1428, \dodoi{10.1093/mnras/stw3237}

\bibitem[{{Guillochon} \& {McCourt}(2017)}]{GuillochonMcCourt2017}
{Guillochon}, J., \& {McCourt}, M. 2017, \apjl, 834, L19, \dodoi{10.3847/2041-8213/834/2/L19}

\bibitem[{{Guo} {et~al.}(2022{\natexlab{a}}){Guo}, {Barth}, \& {Wang}}]{Guo22b}
{Guo}, H., {Barth}, A.~J., \& {Wang}, S. 2022{\natexlab{a}}, \apj, 940, 20, \dodoi{10.3847/1538-4357/ac96ec}

\bibitem[{{Guo} {et~al.}(2022{\natexlab{b}}){Guo}, {Barth}, {Korista}, {Goad}, {Cackett}, {Bentz}, {Brandt}, {Gonzalez-Buitrago}, {Ferland}, {Gelbord}, {Ho}, {Horne}, {Joner}, {Kriss}, {McHardy}, {Mehdipour}, {Park}, {Remigio}, {U}, \& {Vestergaard}}]{Guo22a}
{Guo}, H., {Barth}, A.~J., {Korista}, K.~T., {et~al.} 2022{\natexlab{b}}, \apj, 927, 60, \dodoi{10.3847/1538-4357/ac4bc6}

\bibitem[{{Guolo} {et~al.}(2024){Guolo}, {Gezari}, {Yao}, {van Velzen}, {Hammerstein}, {Cenko}, \& {Tokayer}}]{Guolo24}
{Guolo}, M., {Gezari}, S., {Yao}, Y., {et~al.} 2024, \apj, 966, 160, \dodoi{10.3847/1538-4357/ad2f9f}

\bibitem[{{Hammerstein} {et~al.}(2023){Hammerstein}, {van Velzen}, {Gezari}, {Cenko}, {Yao}, {Ward}, {Frederick}, {Villanueva}, {Somalwar}, {Graham}, {Kulkarni}, {Stern}, {Andreoni}, {Bellm}, {Dekany}, {Dhawan}, {Drake}, {Fremling}, {Gatkine}, {Groom}, {Ho}, {Kasliwal}, {Karambelkar}, {Kool}, {Masci}, {Medford}, {Perley}, {Purdum}, {van Roestel}, {Sharma}, {Sollerman}, {Taggart}, \& {Yan}}]{Hammerstein23}
{Hammerstein}, E., {van Velzen}, S., {Gezari}, S., {et~al.} 2023, \apj, 942, 9, \dodoi{10.3847/1538-4357/aca283}

\bibitem[{{Hern{\'a}ndez Santisteban} {et~al.}(2020){Hern{\'a}ndez Santisteban}, {Edelson}, {Horne}, {Gelbord}, {Barth}, {Cackett}, {Goad}, {Netzer}, {Starkey}, {Uttley}, {Brandt}, {Korista}, {Lohfink}, {Onken}, {Page}, {Siegel}, {Vestergaard}, {Bisogni}, {Breeveld}, {Cenko}, {Dalla Bont{\`a}}, {Evans}, {Ferland}, {Gonzalez-Buitrago}, {Grupe}, {Joner}, {Kriss}, {LaPorte}, {Mathur}, {Marshall}, {Mehdipour}, {Mudd}, {Peterson}, {Schmidt}, {Vaughan}, \& {Valenti}}]{HS20}
{Hern{\'a}ndez Santisteban}, J.~V., {Edelson}, R., {Horne}, K., {et~al.} 2020, \mnras, 498, 5399, \dodoi{10.1093/mnras/staa2365}

\bibitem[{{Hinkle} {et~al.}(2021){Hinkle}, {Holoien}, {Auchettl}, {Shappee}, {Neustadt}, {Payne}, {Brown}, {Kochanek}, {Stanek}, {Graham}, {Tucker}, {Do}, {Anderson}, {Bose}, {Chen}, {Coulter}, {Dimitriadis}, {Dong}, {Foley}, {Huber}, {Hung}, {Kilpatrick}, {Pignata}, {Piro}, {Rojas-Bravo}, {Siebert}, {Stalder}, {Thompson}, {Tonry}, {Vallely}, \& {Wisniewski}}]{Hinkle21}
{Hinkle}, J.~T., {Holoien}, T.~W.~S., {Auchettl}, K., {et~al.} 2021, \mnras, 500, 1673, \dodoi{10.1093/mnras/staa3170}

\bibitem[{{Holoien} {et~al.}(2016){Holoien}, {Kochanek}, {Prieto}, {Grupe}, {Chen}, {Godoy-Rivera}, {Stanek}, {Shappee}, {Dong}, {Brown}, {Basu}, {Beacom}, {Bersier}, {Brimacombe}, {Carlson}, {Falco}, {Johnston}, {Madore}, {Pojmanski}, \& {Seibert}}]{Holoien16}
{Holoien}, T.~W.~S., {Kochanek}, C.~S., {Prieto}, J.~L., {et~al.} 2016, \mnras, 463, 3813, \dodoi{10.1093/mnras/stw2272}

\bibitem[{{Huang} {et~al.}(2023{\natexlab{a}}){Huang}, {Jiang}, {Feng}, {Davis}, {Stone}, \& {Middleton}}]{Huang+2023}
{Huang}, J., {Jiang}, Y.-F., {Feng}, H., {et~al.} 2023{\natexlab{a}}, \apj, 945, 57, \dodoi{10.3847/1538-4357/acb6fc}

\bibitem[{{Huang} {et~al.}(2024){Huang}, {Jiang}, {Zhu}, {Wang}, {Wang}, {Wang}, {Gan}, {Liang}, {Qin}, {Lin}, {Xu}, {Cai}, {Jiang}, {Kong}, {Li}, {li}, {Wang}, {Xu}, {Xue}, {Yuan}, {Cheng}, {Fan}, {Gao}, {Hu}, {Hu}, {Li}, {Li}, {Liang}, {Liu}, {Liu}, {Lou}, {Luo}, {Qian}, {Tang}, {Wan}, {Wang}, {Wang}, {Yang}, {Yao}, {Zhang}, {Zhang}, {Zhao}, {Zheng}, {Zhu}, \& {Zuo}}]{Huang24}
{Huang}, S., {Jiang}, N., {Zhu}, J., {et~al.} 2024, \apjl, 964, L22, \dodoi{10.3847/2041-8213/ad319f}

\bibitem[{{Huang} {et~al.}(2023{\natexlab{b}}){Huang}, {Davis}, \& {Jiang}}]{Huang23}
{Huang}, X., {Davis}, S.~W., \& {Jiang}, Y.-f. 2023{\natexlab{b}}, \apj, 953, 117, \dodoi{10.3847/1538-4357/ace0be}

\bibitem[{{Hung} {et~al.}(2021){Hung}, {Foley}, {Veilleux}, {Cenko}, {Dai}, {Auchettl}, {Brink}, {Dimitriadis}, {Filippenko}, {Gezari}, {Holoien}, {Kilpatrick}, {Mockler}, {Piro}, {Ramirez-Ruiz}, {Rojas-Bravo}, {Siebert}, {van Velzen}, \& {Zheng}}]{Hung21}
{Hung}, T., {Foley}, R.~J., {Veilleux}, S., {et~al.} 2021, \apj, 917, 9, \dodoi{10.3847/1538-4357/abf4c3}

\bibitem[{{Ivezi{\'c}} {et~al.}(2019){Ivezi{\'c}}, {Kahn}, {Tyson}, {Abel}, {Acosta}, {Allsman}, {Alonso}, {AlSayyad}, {Anderson}, {Andrew}, {Angel}, {Angeli}, {Ansari}, {Antilogus}, {Araujo}, {Armstrong}, {Arndt}, {Astier}, {Aubourg}, {Auza}, {Axelrod}, {Bard}, {Barr}, {Barrau}, {Bartlett}, {Bauer}, {Bauman}, {Baumont}, {Bechtol}, {Bechtol}, {Becker}, {Becla}, {Beldica}, {Bellavia}, {Bianco}, {Biswas}, {Blanc}, {Blazek}, {Blandford}, {Bloom}, {Bogart}, {Bond}, {Booth}, {Borgland}, {Borne}, {Bosch}, {Boutigny}, {Brackett}, {Bradshaw}, {Brandt}, {Brown}, {Bullock}, {Burchat}, {Burke}, {Cagnoli}, {Calabrese}, {Callahan}, {Callen}, {Carlin}, {Carlson}, {Chandrasekharan}, {Charles-Emerson}, {Chesley}, {Cheu}, {Chiang}, {Chiang}, {Chirino}, {Chow}, {Ciardi}, {Claver}, {Cohen-Tanugi}, {Cockrum}, {Coles}, {Connolly}, {Cook}, {Cooray}, {Covey}, {Cribbs}, {Cui}, {Cutri}, {Daly}, {Daniel}, {Daruich}, {Daubard}, {Daues}, {Dawson}, {Delgado}, {Dellapenna}, {de Peyster}, {de Val-Borro}, {Digel}, {Doherty}, {Dubois},
  {Dubois-Felsmann}, {Durech}, {Economou}, {Eifler}, {Eracleous}, {Emmons}, {Fausti Neto}, {Ferguson}, {Figueroa}, {Fisher-Levine}, {Focke}, {Foss}, {Frank}, {Freemon}, {Gangler}, {Gawiser}, {Geary}, {Gee}, {Geha}, {Gessner}, {Gibson}, {Gilmore}, {Glanzman}, {Glick}, {Goldina}, {Goldstein}, {Goodenow}, {Graham}, {Gressler}, {Gris}, {Guy}, {Guyonnet}, {Haller}, {Harris}, {Hascall}, {Haupt}, {Hernandez}, {Herrmann}, {Hileman}, {Hoblitt}, {Hodgson}, {Hogan}, {Howard}, {Huang}, {Huffer}, {Ingraham}, {Innes}, {Jacoby}, {Jain}, {Jammes}, {Jee}, {Jenness}, {Jernigan}, {Jevremovi{\'c}}, {Johns}, {Johnson}, {Johnson}, {Jones}, {Juramy-Gilles}, {Juri{\'c}}, {Kalirai}, {Kallivayalil}, {Kalmbach}, {Kantor}, {Karst}, {Kasliwal}, {Kelly}, {Kessler}, {Kinnison}, {Kirkby}, {Knox}, {Kotov}, {Krabbendam}, {Krughoff}, {Kub{\'a}nek}, {Kuczewski}, {Kulkarni}, {Ku}, {Kurita}, {Lage}, {Lambert}, {Lange}, {Langton}, {Le Guillou}, {Levine}, {Liang}, {Lim}, {Lintott}, {Long}, {Lopez}, {Lotz}, {Lupton}, {Lust}, {MacArthur}, {Mahabal},
  {Mandelbaum}, {Markiewicz}, {Marsh}, {Marshall}, {Marshall}, {May}, {McKercher}, {McQueen}, {Meyers}, {Migliore}, {Miller}, {Mills}, {Miraval}, {Moeyens}, {Moolekamp}, {Monet}, {Moniez}, {Monkewitz}, {Montgomery}, {Morrison}, {Mueller}, {Muller}, {Mu{\~n}oz Arancibia}, {Neill}, {Newbry}, {Nief}, {Nomerotski}, {Nordby}, {O'Connor}, {Oliver}, {Olivier}, {Olsen}, {O'Mullane}, {Ortiz}, {Osier}, {Owen}, {Pain}, {Palecek}, {Parejko}, {Parsons}, {Pease}, {Peterson}, {Peterson}, {Petravick}, {Libby Petrick}, {Petry}, {Pierfederici}, {Pietrowicz}, {Pike}, {Pinto}, {Plante}, {Plate}, {Plutchak}, {Price}, {Prouza}, {Radeka}, {Rajagopal}, {Rasmussen}, {Regnault}, {Reil}, {Reiss}, {Reuter}, {Ridgway}, {Riot}, {Ritz}, {Robinson}, {Roby}, {Roodman}, {Rosing}, {Roucelle}, {Rumore}, {Russo}, {Saha}, {Sassolas}, {Schalk}, {Schellart}, {Schindler}, {Schmidt}, {Schneider}, {Schneider}, {Schoening}, {Schumacher}, {Schwamb}, {Sebag}, {Selvy}, {Sembroski}, {Seppala}, {Serio}, {Serrano}, {Shaw}, {Shipsey}, {Sick}, {Silvestri},
  {Slater}, {Smith}, {Smith}, {Sobhani}, {Soldahl}, {Storrie-Lombardi}, {Stover}, {Strauss}, {Street}, {Stubbs}, {Sullivan}, {Sweeney}, {Swinbank}, {Szalay}, {Takacs}, {Tether}, {Thaler}, {Thayer}, {Thomas}, {Thornton}, {Thukral}, {Tice}, {Trilling}, {Turri}, {Van Berg}, {Vanden Berk}, {Vetter}, {Virieux}, {Vucina}, {Wahl}, {Walkowicz}, {Walsh}, {Walter}, {Wang}, {Wang}, {Warner}, {Wiecha}, {Willman}, {Winters}, {Wittman}, {Wolff}, {Wood-Vasey}, {Wu}, {Xin}, {Yoachim}, \& {Zhan}}]{Zeljko19}
{Ivezi{\'c}}, {\v{Z}}., {Kahn}, S.~M., {Tyson}, J.~A., {et~al.} 2019, \apj, 873, 111, \dodoi{10.3847/1538-4357/ab042c}

\bibitem[{{Jiang} {et~al.}(2021){Jiang}, {Wang}, {Hu}, {Sun}, {Dou}, \& {Xiao}}]{Jiang21a}
{Jiang}, N., {Wang}, T., {Hu}, X., {et~al.} 2021, \apj, 911, 31, \dodoi{10.3847/1538-4357/abe772}

\bibitem[{{Jiang} {et~al.}(2019){Jiang}, {Blaes}, {Stone}, \& {Davis}}]{Jiang19+Corona}
{Jiang}, Y.-F., {Blaes}, O., {Stone}, J.~M., \& {Davis}, S.~W. 2019, \apj, 885, 144, \dodoi{10.3847/1538-4357/ab4a00}

\bibitem[{{Jiang} {et~al.}(2016){Jiang}, {Guillochon}, \& {Loeb}}]{Jiang16}
{Jiang}, Y.-F., {Guillochon}, J., \& {Loeb}, A. 2016, \apj, 830, 125, \dodoi{10.3847/0004-637X/830/2/125}

\bibitem[{{Jonker} {et~al.}(2020){Jonker}, {Stone}, {Generozov}, {van Velzen}, \& {Metzger}}]{Jonker20}
{Jonker}, P.~G., {Stone}, N.~C., {Generozov}, A., {van Velzen}, S., \& {Metzger}, B. 2020, \apj, 889, 166, \dodoi{10.3847/1538-4357/ab659c}

\bibitem[{{Kara} {et~al.}(2021){Kara}, {Mehdipour}, {Kriss}, {Cackett}, {Arav}, {Barth}, {Byun}, {Brotherton}, {De Rosa}, {Gelbord}, {Hern{\'a}ndez Santisteban}, {Hu}, {Kaastra}, {Landt}, {Li}, {Miller}, {Montano}, {Partington}, {Aceituno}, {Bai}, {Bao}, {Bentz}, {Brink}, {Chelouche}, {Chen}, {Colmenero}, {Dalla Bont{\`a}}, {Dehghanian}, {Du}, {Edelson}, {Ferland}, {Ferrarese}, {Fian}, {Filippenko}, {Fischer}, {Goad}, {Gonz{\'a}lez Buitrago}, {Gorjian}, {Grier}, {Guo}, {Hall}, {Ho}, {Homayouni}, {Horne}, {Ili{\'c}}, {Jiang}, {Joner}, {Kaspi}, {Kochanek}, {Korista}, {Kynoch}, {Li}, {Liu}, {McHardy}, {McLane}, {Mitchell}, {Netzer}, {Olson}, {Pogge}, {Popovi{\'c}}, {Proga}, {Storchi-Bergmann}, {Strasburger}, {Treu}, {Vestergaard}, {Wang}, {Ward}, {Waters}, {Williams}, {Yang}, {Yao}, {Zastrocky}, {Zhai}, \& {Zu}}]{Kara21}
{Kara}, E., {Mehdipour}, M., {Kriss}, G.~A., {et~al.} 2021, \apj, 922, 151, \dodoi{10.3847/1538-4357/ac2159}

\bibitem[{{Komossa} \& {Bade}(1999)}]{Komossa99}
{Komossa}, S., \& {Bade}, N. 1999, \aap, 343, 775.
\newblock \doarXiv{astro-ph/9901141}

\bibitem[{{Koratkar} \& {Blaes}(1999)}]{KoratkarBlaes1999}
{Koratkar}, A., \& {Blaes}, O. 1999, \pasp, 111, 1, \dodoi{10.1086/316294}

\bibitem[{{Kr{\"u}hler} {et~al.}(2018){Kr{\"u}hler}, {Fraser}, {Leloudas}, {Schulze}, {Stone}, {van Velzen}, {Amorin}, {Hjorth}, {Jonker}, {Kann}, {Kim}, {Kuncarayakti}, {Mehner}, \& {Nicuesa Guelbenzu}}]{Kruhler18}
{Kr{\"u}hler}, T., {Fraser}, M., {Leloudas}, G., {et~al.} 2018, \aap, 610, A14, \dodoi{10.1051/0004-6361/201731773}

\bibitem[{{Leloudas} {et~al.}(2016){Leloudas}, {Fraser}, {Stone}, {van Velzen}, {Jonker}, {Arcavi}, {Fremling}, {Maund}, {Smartt}, {Kr{\`\i}hler}, {Miller-Jones}, {Vreeswijk}, {Gal-Yam}, {Mazzali}, {De Cia}, {Howell}, {Inserra}, {Patat}, {de Ugarte Postigo}, {Yaron}, {Ashall}, {Bar}, {Campbell}, {Chen}, {Childress}, {Elias-Rosa}, {Harmanen}, {Hosseinzadeh}, {Johansson}, {Kangas}, {Kankare}, {Kim}, {Kuncarayakti}, {Lyman}, {Magee}, {Maguire}, {Malesani}, {Mattila}, {McCully}, {Nicholl}, {Prentice}, {Romero-Ca{\~n}izales}, {Schulze}, {Smith}, {Sollerman}, {Sullivan}, {Tucker}, {Valenti}, {Wheeler}, \& {Young}}]{Leloudas16}
{Leloudas}, G., {Fraser}, M., {Stone}, N.~C., {et~al.} 2016, Nature Astronomy, 1, 0002, \dodoi{10.1038/s41550-016-0002}

\bibitem[{{Lin} {et~al.}(2024){Lin}, {Jiang}, {Wang}, {Kong}, {Li}, {He}, {Wang}, {Zhu}, {Li}, {Jiang}, {Singh}, {Singh Teja}, {Sahu}, {Jin}, {Maeda}, \& {Huang}}]{Lin24}
{Lin}, Z., {Jiang}, N., {Wang}, T., {et~al.} 2024, arXiv e-prints, arXiv:2405.10895, \dodoi{10.48550/arXiv.2405.10895}

\bibitem[{{Liodakis} {et~al.}(2023){Liodakis}, {Koljonen}, {Blinov}, {Lindfors}, {Alexander}, {Hovatta}, {Berton}, {Hajela}, {Jormanainen}, {Kouroumpatzakis}, {Mandarakas}, \& {Nilsson}}]{Liodakis2023}
{Liodakis}, I., {Koljonen}, K.~I.~I., {Blinov}, D., {et~al.} 2023, Science, 380, 656, \dodoi{10.1126/science.abj9570}

\bibitem[{{Liu} {et~al.}(2022){Liu}, {Dou}, {Chen}, \& {Shen}}]{Liu22}
{Liu}, X.-L., {Dou}, L.-M., {Chen}, J.-H., \& {Shen}, R.-F. 2022, \apj, 925, 67, \dodoi{10.3847/1538-4357/ac33a9}

\bibitem[{{Liu} {et~al.}(2023){Liu}, {Malyali}, {Krumpe}, {Homan}, {Goodwin}, {Grotova}, {Kawka}, {Rau}, {Merloni}, {Anderson}, {Miller-Jones}, {Markowitz}, {Ciroi}, {Di Mille}, {Schramm}, {Tang}, {Buckley}, {Gromadzki}, {Jin}, \& {Buchner}}]{Liu23}
{Liu}, Z., {Malyali}, A., {Krumpe}, M., {et~al.} 2023, \aap, 669, A75, \dodoi{10.1051/0004-6361/202244805}

\bibitem[{{Lobban} {et~al.}(2020){Lobban}, {Zola}, {Pajdosz-{\'S}mierciak}, {Braito}, {Nardini}, {Bhatta}, {Markowitz}, {Bachev}, {Carosati}, {Caton}, {Damljanovic}, {D{\k{e}}bski}, {Haislip}, {Hu}, {Kouprianov}, {Krzesi{\'n}ski}, {Porquet}, {Pozo Nu{\~n}ez}, {Reeves}, \& {Reichart}}]{Lobban20}
{Lobban}, A.~P., {Zola}, S., {Pajdosz-{\'S}mierciak}, U., {et~al.} 2020, \mnras, 494, 1165, \dodoi{10.1093/mnras/staa630}

\bibitem[{{Loeb} \& {Ulmer}(1997)}]{Loeb97}
{Loeb}, A., \& {Ulmer}, A. 1997, \apj, 489, 573, \dodoi{10.1086/304814}

\bibitem[{{Lu} \& {Bonnerot}(2020)}]{Lu20}
{Lu}, W., \& {Bonnerot}, C. 2020, \mnras, 492, 686, \dodoi{10.1093/mnras/stz3405}

\bibitem[{{Lu} \& {Kumar}(2018)}]{Lu18}
{Lu}, W., \& {Kumar}, P. 2018, \apj, 865, 128, \dodoi{10.3847/1538-4357/aad54a}

\bibitem[{{Malyali} {et~al.}(2021){Malyali}, {Rau}, {Merloni}, {Nandra}, {Buchner}, {Liu}, {Gezari}, {Sollerman}, {Shappee}, {Trakhtenbrot}, {Arcavi}, {Ricci}, {van Velzen}, {Goobar}, {Frederick}, {Kawka}, {Tartaglia}, {Burke}, {Hiramatsu}, {Schramm}, {van der Boom}, {Anderson}, {Miller-Jones}, {Bellm}, {Drake}, {Duev}, {Fremling}, {Graham}, {Masci}, {Rusholme}, {Soumagnac}, \& {Walters}}]{Malyali21}
{Malyali}, A., {Rau}, A., {Merloni}, A., {et~al.} 2021, \aap, 647, A9, \dodoi{10.1051/0004-6361/202039681}

\bibitem[{{Margutti} {et~al.}(2017){Margutti}, {Metzger}, {Chornock}, {Milisavljevic}, {Berger}, {Blanchard}, {Guidorzi}, {Migliori}, {Kamble}, {Lunnan}, {Nicholl}, {Coppejans}, {Dall'Osso}, {Drout}, {Perna}, \& {Sbarufatti}}]{Margutti17}
{Margutti}, R., {Metzger}, B.~D., {Chornock}, R., {et~al.} 2017, \apj, 836, 25, \dodoi{10.3847/1538-4357/836/1/25}

\bibitem[{{Masterson} {et~al.}(2024){Masterson}, {De}, {Panagiotou}, {Kara}, {Arcavi}, {Eilers}, {Frostig}, {Gezari}, {Grotova}, {Liu}, {Malyali}, {Meisner}, {Merloni}, {Newsome}, {Rau}, {Simcoe}, \& {van Velzen}}]{Masterson24}
{Masterson}, M., {De}, K., {Panagiotou}, C., {et~al.} 2024, \apj, 961, 211, \dodoi{10.3847/1538-4357/ad18bb}

\bibitem[{{McHardy} {et~al.}(2014){McHardy}, {Cameron}, {Dwelly}, {Connolly}, {Lira}, {Emmanoulopoulos}, {Gelbord}, {Breedt}, {Arevalo}, \& {Uttley}}]{McHardy14}
{McHardy}, I.~M., {Cameron}, D.~T., {Dwelly}, T., {et~al.} 2014, \mnras, 444, 1469, \dodoi{10.1093/mnras/stu1636}

\bibitem[{{Metzger}(2022)}]{Metzger22}
{Metzger}, B.~D. 2022, \apjl, 937, L12, \dodoi{10.3847/2041-8213/ac90ba}

\bibitem[{{Metzger} \& {Stone}(2016)}]{Metzger16}
{Metzger}, B.~D., \& {Stone}, N.~C. 2016, \mnras, 461, 948, \dodoi{10.1093/mnras/stw1394}

\bibitem[{{Mishra} {et~al.}(2022){Mishra}, {Fragile}, {Anderson}, {Blankenship}, {Li}, \& {Nalewajko}}]{Mishra+2022}
{Mishra}, B., {Fragile}, P.~C., {Anderson}, J., {et~al.} 2022, \apj, 939, 31, \dodoi{10.3847/1538-4357/ac938b}

\bibitem[{{Mummery} \& {Balbus}(2020)}]{Mummery20}
{Mummery}, A., \& {Balbus}, S.~A. 2020, \mnras, 497, L13, \dodoi{10.1093/mnrasl/slaa105}

\bibitem[{{Nicholl} {et~al.}(2020){Nicholl}, {Wevers}, {Oates}, {Alexander}, {Leloudas}, {Onori}, {Jerkstrand}, {Gomez}, {Campana}, {Arcavi}, {Charalampopoulos}, {Gromadzki}, {Ihanec}, {Jonker}, {Lawrence}, {Mandel}, {Schulze}, {Short}, {Burke}, {McCully}, {Hiramatsu}, {Howell}, {Pellegrino}, {Abbot}, {Anderson}, {Berger}, {Blanchard}, {Cannizzaro}, {Chen}, {Dennefeld}, {Galbany}, {Gonz{\'a}lez-Gait{\'a}n}, {Hosseinzadeh}, {Inserra}, {Irani}, {Kuin}, {M{\"u}ller-Bravo}, {Pineda}, {Ross}, {Roy}, {Smartt}, {Smith}, {Tucker}, {Wyrzykowski}, \& {Young}}]{Nicholl20}
{Nicholl}, M., {Wevers}, T., {Oates}, S.~R., {et~al.} 2020, \mnras, 499, 482, \dodoi{10.1093/mnras/staa2824}

\bibitem[{{Paczynski}(1983)}]{Paczynski83}
{Paczynski}, B. 1983, \apj, 267, 315, \dodoi{10.1086/160870}

\bibitem[{{Pahari} {et~al.}(2020){Pahari}, {McHardy}, {Vincentelli}, {Cackett}, {Peterson}, {Goad}, {G{\"u}ltekin}, \& {Horne}}]{Pahari20}
{Pahari}, M., {McHardy}, I.~M., {Vincentelli}, F., {et~al.} 2020, \mnras, 494, 4057, \dodoi{10.1093/mnras/staa1055}

\bibitem[{{Parkinson} {et~al.}(2022){Parkinson}, {Knigge}, {Matthews}, {Long}, {Higginbottom}, {Sim}, \& {Mangham}}]{Parkinson22}
{Parkinson}, E.~J., {Knigge}, C., {Matthews}, J.~H., {et~al.} 2022, \mnras, 510, 5426, \dodoi{10.1093/mnras/stac027}

\bibitem[{{Pasham} {et~al.}(2017){Pasham}, {Cenko}, {Sadowski}, {Guillochon}, {Stone}, {van Velzen}, \& {Cannizzo}}]{Pasham17}
{Pasham}, D.~R., {Cenko}, S.~B., {Sadowski}, A., {et~al.} 2017, \apjl, 837, L30, \dodoi{10.3847/2041-8213/aa6003}

\bibitem[{{Payne} {et~al.}(2021){Payne}, {Shappee}, {Hinkle}, {Vallely}, {Kochanek}, {Holoien}, {Auchettl}, {Stanek}, {Thompson}, {Neustadt}, {Tucker}, {Armstrong}, {Brimacombe}, {Cacella}, {Cornect}, {Denneau}, {Fausnaugh}, {Flewelling}, {Grupe}, {Heinze}, {Lopez}, {Monard}, {Prieto}, {Schneider}, {Sheppard}, {Tonry}, \& {Weiland}}]{Payne21}
{Payne}, A.~V., {Shappee}, B.~J., {Hinkle}, J.~T., {et~al.} 2021, \apj, 910, 125, \dodoi{10.3847/1538-4357/abe38d}

\bibitem[{{Peterson} {et~al.}(1998){Peterson}, {Wanders}, {Horne}, {Collier}, {Alexander}, {Kaspi}, \& {Maoz}}]{Peterson98}
{Peterson}, B.~M., {Wanders}, I., {Horne}, K., {et~al.} 1998, \pasp, 110, 660, \dodoi{10.1086/316177}

\bibitem[{{Phinney}(1989)}]{Phinney89}
{Phinney}, E.~S. 1989, in The Center of the Galaxy, ed. M.~{Morris}, Vol. 136, 543

\bibitem[{{Piran} {et~al.}(2015){Piran}, {Svirski}, {Krolik}, {Cheng}, \& {Shiokawa}}]{Piran15}
{Piran}, T., {Svirski}, G., {Krolik}, J., {Cheng}, R.~M., \& {Shiokawa}, H. 2015, \apj, 806, 164, \dodoi{10.1088/0004-637X/806/2/164}

\bibitem[{{Rees}(1988)}]{Rees88}
{Rees}, M.~J. 1988, \nat, 333, 523, \dodoi{10.1038/333523a0}

\bibitem[{{Roth} {et~al.}(2016){Roth}, {Kasen}, {Guillochon}, \& {Ramirez-Ruiz}}]{Roth16}
{Roth}, N., {Kasen}, D., {Guillochon}, J., \& {Ramirez-Ruiz}, E. 2016, \apj, 827, 3, \dodoi{10.3847/0004-637X/827/1/3}

\bibitem[{{Ryu} {et~al.}(2023){Ryu}, {Krolik}, {Piran}, {Noble}, \& {Avara}}]{Ryu23}
{Ryu}, T., {Krolik}, J., {Piran}, T., {Noble}, S.~C., \& {Avara}, M. 2023, \apj, 957, 12, \dodoi{10.3847/1538-4357/acf5de}

\bibitem[{{Sazonov} {et~al.}(2021){Sazonov}, {Gilfanov}, {Medvedev}, {Yao}, {Khorunzhev}, {Semena}, {Sunyaev}, {Burenin}, {Lyapin}, {Meshcheryakov}, {Uskov}, {Zaznobin}, {Postnov}, {Dodin}, {Belinski}, {Cherepashchuk}, {Eselevich}, {Dodonov}, {Grokhovskaya}, {Kotov}, {Bikmaev}, {Zhuchkov}, {Gumerov}, {van Velzen}, \& {Kulkarni}}]{Sazonov21}
{Sazonov}, S., {Gilfanov}, M., {Medvedev}, P., {et~al.} 2021, \mnras, 508, 3820, \dodoi{10.1093/mnras/stab2843}

\bibitem[{{Shakura} \& {Sunyaev}(1973)}]{Shakura73}
{Shakura}, N.~I., \& {Sunyaev}, R.~A. 1973, \aap, 24, 337

\bibitem[{{Shappee} {et~al.}(2014){Shappee}, {Prieto}, {Grupe}, {Kochanek}, {Stanek}, {De Rosa}, {Mathur}, {Zu}, {Peterson}, {Pogge}, {Komossa}, {Im}, {Jencson}, {Holoien}, {Basu}, {Beacom}, {Szczygie{\l}}, {Brimacombe}, {Adams}, {Campillay}, {Choi}, {Contreras}, {Dietrich}, {Dubberley}, {Elphick}, {Foale}, {Giustini}, {Gonzalez}, {Hawkins}, {Howell}, {Hsiao}, {Koss}, {Leighly}, {Morrell}, {Mudd}, {Mullins}, {Nugent}, {Parrent}, {Phillips}, {Pojmanski}, {Rosing}, {Ross}, {Sand}, {Terndrup}, {Valenti}, {Walker}, \& {Yoon}}]{Shappee14}
{Shappee}, B.~J., {Prieto}, J.~L., {Grupe}, D., {et~al.} 2014, \apj, 788, 48, \dodoi{10.1088/0004-637X/788/1/48}

\bibitem[{{Shen} \& {Matzner}(2014)}]{Shen14}
{Shen}, R.-F., \& {Matzner}, C.~D. 2014, \apj, 784, 87, \dodoi{10.1088/0004-637X/784/2/87}

\bibitem[{{Shiokawa} {et~al.}(2015){Shiokawa}, {Krolik}, {Cheng}, {Piran}, \& {Noble}}]{Shiokawa15}
{Shiokawa}, H., {Krolik}, J.~H., {Cheng}, R.~M., {Piran}, T., \& {Noble}, S.~C. 2015, \apj, 804, 85, \dodoi{10.1088/0004-637X/804/2/85}

\bibitem[{{Somalwar} {et~al.}(2023){Somalwar}, {Ravi}, {Yao}, {Guolo}, {Graham}, {Hammerstein}, {Lu}, {Nicholl}, {Sharma}, {Stein}, {van Velzen}, {Bellm}, {Coughlin}, {Groom}, {Masci}, \& {Riddle}}]{Somalwar23}
{Somalwar}, J.~J., {Ravi}, V., {Yao}, Y., {et~al.} 2023, arXiv e-prints, arXiv:2310.03782, \dodoi{10.48550/arXiv.2310.03782}

\bibitem[{{Stein} {et~al.}(2021){Stein}, {van Velzen}, {Kowalski}, {Franckowiak}, {Gezari}, {Miller-Jones}, {Frederick}, {Sfaradi}, {Bietenholz}, {Horesh}, {Fender}, {Garrappa}, {Ahumada}, {Andreoni}, {Belicki}, {Bellm}, {B{\"o}ttcher}, {Brinnel}, {Burruss}, {Cenko}, {Coughlin}, {Cunningham}, {Drake}, {Farrar}, {Feeney}, {Foley}, {Gal-Yam}, {Golkhou}, {Goobar}, {Graham}, {Hammerstein}, {Helou}, {Hung}, {Kasliwal}, {Kilpatrick}, {Kong}, {Kupfer}, {Laher}, {Mahabal}, {Masci}, {Necker}, {Nordin}, {Perley}, {Rigault}, {Reusch}, {Rodriguez}, {Rojas-Bravo}, {Rusholme}, {Shupe}, {Singer}, {Sollerman}, {Soumagnac}, {Stern}, {Taggart}, {van Santen}, {Ward}, {Woudt}, \& {Yao}}]{Stein21}
{Stein}, R., {van Velzen}, S., {Kowalski}, M., {et~al.} 2021, Nature Astronomy, 5, 510, \dodoi{10.1038/s41550-020-01295-8}

\bibitem[{{Stone} \& {Metzger}(2016)}]{Stone16}
{Stone}, N.~C., \& {Metzger}, B.~D. 2016, \mnras, 455, 859, \dodoi{10.1093/mnras/stv2281}

\bibitem[{{Strubbe} \& {Quataert}(2009)}]{Strubbe09}
{Strubbe}, L.~E., \& {Quataert}, E. 2009, \mnras, 400, 2070, \dodoi{10.1111/j.1365-2966.2009.15599.x}

\bibitem[{{Sun} {et~al.}(2018){Sun}, {Grier}, \& {Peterson}}]{Sun18}
{Sun}, M., {Grier}, C.~J., \& {Peterson}, B.~M. 2018, {PyCCF: Python Cross Correlation Function for reverberation mapping studies}, Astrophysics Source Code Library, record ascl:1805.032.
\newblock \doeprint{1805.032}

\bibitem[{{Thomsen} {et~al.}(2022){Thomsen}, {Kwan}, {Dai}, {Wu}, {Roth}, \& {Ramirez-Ruiz}}]{Thomsen22}
{Thomsen}, L.~L., {Kwan}, T.~M., {Dai}, L., {et~al.} 2022, \apjl, 937, L28, \dodoi{10.3847/2041-8213/ac911f}

\bibitem[{{van Velzen} {et~al.}(2019){van Velzen}, {Gezari}, {Cenko}, {Kara}, {Miller-Jones}, {Hung}, {Bright}, {Roth}, {Blagorodnova}, {Huppenkothen}, {Yan}, {Ofek}, {Sollerman}, {Frederick}, {Ward}, {Graham}, {Fender}, {Kasliwal}, {Canella}, {Stein}, {Giomi}, {Brinnel}, {van Santen}, {Nordin}, {Bellm}, {Dekany}, {Fremling}, {Golkhou}, {Kupfer}, {Kulkarni}, {Laher}, {Mahabal}, {Masci}, {Miller}, {Neill}, {Riddle}, {Rigault}, {Rusholme}, {Soumagnac}, \& {Tachibana}}]{Vanvelzen19b}
{van Velzen}, S., {Gezari}, S., {Cenko}, S.~B., {et~al.} 2019, \apj, 872, 198, \dodoi{10.3847/1538-4357/aafe0c}

\bibitem[{{van Velzen} {et~al.}(2021){van Velzen}, {Gezari}, {Hammerstein}, {Roth}, {Frederick}, {Ward}, {Hung}, {Cenko}, {Stein}, {Perley}, {Taggart}, {Foley}, {Sollerman}, {Blagorodnova}, {Andreoni}, {Bellm}, {Brinnel}, {De}, {Dekany}, {Feeney}, {Fremling}, {Giomi}, {Golkhou}, {Graham}, {Ho}, {Kasliwal}, {Kilpatrick}, {Kulkarni}, {Kupfer}, {Laher}, {Mahabal}, {Masci}, {Miller}, {Nordin}, {Riddle}, {Rusholme}, {van Santen}, {Sharma}, {Shupe}, \& {Soumagnac}}]{vanVelzen21b}
{van Velzen}, S., {Gezari}, S., {Hammerstein}, E., {et~al.} 2021, \apj, 908, 4, \dodoi{10.3847/1538-4357/abc258}

\bibitem[{{Vincentelli} {et~al.}(2021){Vincentelli}, {McHardy}, {Cackett}, {Barth}, {Horne}, {Goad}, {Korista}, {Gelbord}, {Brandt}, {Edelson}, {Miller}, {Pahari}, {Peterson}, {Schmidt}, {Baldi}, {Breedt}, {Hern{\'a}ndez Santisteban}, {Romero-Colmenero}, {Ward}, \& {Williams}}]{Vincentelli21}
{Vincentelli}, F.~M., {McHardy}, I., {Cackett}, E.~M., {et~al.} 2021, \mnras, 504, 4337, \dodoi{10.1093/mnras/stab1033}

\bibitem[{{Wang} {et~al.}(2023{\natexlab{a}}){Wang}, {Liu}, {Cai}, {Geng}, {Fang}, {He}, {Jiang}, {Jiang}, {Kong}, {Li}, {Li}, {Luo}, {Pan}, {Wu}, {Yang}, {Yu}, {Zheng}, {Zhu}, {Cai}, {Chen}, {Chen}, {Dai}, {Fan}, {Fan}, {Fang}, {He}, {Hu}, {Hu}, {Jin}, {Jiang}, {Li}, {Li}, {Li}, {Liang}, {Lin}, {Liu}, {Liu}, {Liu}, {Liu}, {Liu}, {Lou}, {Qu}, {Sheng}, {Shi}, {Shu}, {Su}, {Sun}, {Wang}, {Wang}, {Wang}, {Wang}, {Wei}, {Wei}, {Xue}, {Yan}, {Yang}, {Yuan}, {Yuan}, {Zhang}, {Zhang}, {Zhao}, \& {Zhao}}]{WangTG23}
{Wang}, T., {Liu}, G., {Cai}, Z., {et~al.} 2023{\natexlab{a}}, Science China Physics, Mechanics, and Astronomy, 66, 109512, \dodoi{10.1007/s11433-023-2197-5}

\bibitem[{{Wang} {et~al.}(2023{\natexlab{b}}){Wang}, {Baldi}, {del Palacio}, {Guolo}, {Yang}, {Zhang}, {Done}, {Castro Segura}, {Pasham}, {Middleton}, {Altamirano}, {Gandhi}, {Qiao}, {Jiang}, {Yan}, {Giroletti}, {Migliori}, {McHardy}, {Panessa}, {Jin}, {Shen}, \& {Dai}}]{Wang23}
{Wang}, Y., {Baldi}, R.~D., {del Palacio}, S., {et~al.} 2023{\natexlab{b}}, \mnras, 520, 2417, \dodoi{10.1093/mnras/stad101}

\bibitem[{{Welsh}(1999)}]{Welsh99}
{Welsh}, W.~F. 1999, \pasp, 111, 1347, \dodoi{10.1086/316457}

\bibitem[{{Wevers}(2020)}]{Wevers20}
{Wevers}, T. 2020, \mnras, 497, L1, \dodoi{10.1093/mnrasl/slaa097}

\bibitem[{{Wevers} {et~al.}(2019){Wevers}, {Pasham}, {van Velzen}, {Leloudas}, {Schulze}, {Miller-Jones}, {Jonker}, {Gromadzki}, {Kankare}, {Hodgkin}, {Wyrzykowski}, {Kostrzewa-Rutkowska}, {Moran}, {Berton}, {Maguire}, {Onori}, {Mattila}, \& {Nicholl}}]{Wevers19a}
{Wevers}, T., {Pasham}, D.~R., {van Velzen}, S., {et~al.} 2019, \mnras, 488, 4816, \dodoi{10.1093/mnras/stz1976}

\bibitem[{{Wevers} {et~al.}(2021){Wevers}, {Pasham}, {van Velzen}, {Miller-Jones}, {Uttley}, {Gendreau}, {Remillard}, {Arzoumanian}, {L{\"o}wenstein}, \& {Chiti}}]{Wevers21}
---. 2021, \apj, 912, 151, \dodoi{10.3847/1538-4357/abf5e2}

\bibitem[{{Wevers} {et~al.}(2023){Wevers}, {Coughlin}, {Pasham}, {Guolo}, {Sun}, {Wen}, {Jonker}, {Zabludoff}, {Malyali}, {Arcodia}, {Liu}, {Merloni}, {Rau}, {Grotova}, {Short}, \& {Cao}}]{Wevers23}
{Wevers}, T., {Coughlin}, E.~R., {Pasham}, D.~R., {et~al.} 2023, \apjl, 942, L33, \dodoi{10.3847/2041-8213/ac9f36}

\bibitem[{{White} \& {Peterson}(1994)}]{White94}
{White}, R.~J., \& {Peterson}, B.~M. 1994, \pasp, 106, 879, \dodoi{10.1086/133456}

\bibitem[{{Wiseman} {et~al.}(2024){Wiseman}, {Williams}, {Arcavi}, {Galbany}, {Graham}, {H{\"o}nig}, {Newsome}, {Subrayan}, {Sullivan}, {Wang}, {Ili{\'c}}, {Nicholl}, {Oates}, {Petrushevska}, \& {Smith}}]{Wiseman24}
{Wiseman}, P., {Williams}, R.~D., {Arcavi}, I., {et~al.} 2024, arXiv e-prints, arXiv:2406.11552, \dodoi{10.48550/arXiv.2406.11552}

\bibitem[{{Yao} {et~al.}(2022){Yao}, {Lu}, {Guolo}, {Pasham}, {Gezari}, {Gilfanov}, {Gendreau}, {Harrison}, {Cenko}, {Kulkarni}, {Miller}, {Walton}, {Garc{\'\i}a}, {van Velzen}, {Alexander}, {Miller-Jones}, {Nicholl}, {Hammerstein}, {Medvedev}, {Stern}, {Ravi}, {Sunyaev}, {Bloom}, {Graham}, {Kool}, {Mahabal}, {Masci}, {Purdum}, {Rusholme}, {Sharma}, {Smith}, \& {Sollerman}}]{Yao22}
{Yao}, Y., {Lu}, W., {Guolo}, M., {et~al.} 2022, \apj, 937, 8, \dodoi{10.3847/1538-4357/ac898a}

\bibitem[{{Yao} {et~al.}(2023){Yao}, {Ravi}, {Gezari}, {van Velzen}, {Lu}, {Schulze}, {Somalwar}, {Kulkarni}, {Hammerstein}, {Nicholl}, {Graham}, {Perley}, {Cenko}, {Stein}, {Ricarte}, {Chadayammuri}, {Quataert}, {Bellm}, {Bloom}, {Dekany}, {Drake}, {Groom}, {Mahabal}, {Prince}, {Riddle}, {Rusholme}, {Sharma}, {Sollerman}, \& {Yan}}]{Yao23}
{Yao}, Y., {Ravi}, V., {Gezari}, S., {et~al.} 2023, \apjl, 955, L6, \dodoi{10.3847/2041-8213/acf216}

\bibitem[{{Yuan} {et~al.}(2022){Yuan}, {Zhang}, {Chen}, \& {Ling}}]{Yuan22}
{Yuan}, W., {Zhang}, C., {Chen}, Y., \& {Ling}, Z. 2022, in Handbook of X-ray and Gamma-ray Astrophysics, 86, \dodoi{10.1007/978-981-16-4544-0_151-1}

\end{thebibliography}
\bibliographystyle{aasjournal}

\end{CJK}

\appendix

\section{Diffuse continuum}\label{app:DC}
CRM of AGN has demonstrated that AGN continuum variations exhibit wavelength-dependent and positive-direction lags that scale as $\tau \propto \lambda^{4/3}$, which is consistent with the prediction of the standard thin disk model \citep{Shakura73}. These inter-band continuum lags are expected within the framework of the X-ray reprocessing model, wherein high-energy photons emitted from the corona above the disk center are absorbed and reprocessed to lower energies at larger radii in the disk. By measuring these lags, which correspond to the light-travel time across the disk, continuum reverberation can yield valuable insights into determine the sizes of AGN accretion disks. However, current AGN CRM campaigns have revealed that the inferred disk sizes, based on the lags, are several times larger than the predictions of the standard disk model \citep{Cackett21}. Robust evidence of the lag excess around Balmer jump at 3647~\AA\ (or $u/U$ band lag excess) relative to the scaling relation ($\tau \propto \lambda^{4/3}$) indicates the unexpected large continuum lags mainly attributed to the diffuse continuum originating from the inner broad-line region \citep{Cackett18}. The mechanism of the diffuse continuum covering from $\sim$ 1000 $-$ 10000 \AA\ is dominated by free-free, free-bound, and scattered continuum emission of Hydrogen and Helium and excited by X-ray/EUV photons \citep{Fausnaugh16,Cackett21,Guo22b,Guo22a}. 

We compared positive-direction continuum lags in three TDEs (AT~2019qiz in Figure \ref{fig:2019qiz}, AT~2019azh and AT~2020zso in Figure \ref{fig:UV_lag}) with that in AGNs, and concluded that the continuum lags in TDE are unlikely caused by diffuse continuum for three reasons: Firstly, limited high-energy photons (e.g., X-ray/EUV photons) during the first peak (episode II in Figure \ref{fig:cartoon}) makes it unlikely to be excited and contribute significantly to the first light peak before the formation of the disk. Note that since the lag of a few days are common in normal TDEs residing in quiescent galaxies, we suggest the pre-exist weak AGN in AT~2019qiz has negligible contribution to the observed continuum lags. Secondly, there is no clear $u/U$-band lag excesses around 3647 \AA\ in the rest frame, as commonly seen in most AGNs. Finally, the best-fit power-law slopes observed in our TDEs is steeper than that from the standard disk and diffuse continuum. Therefore, we suggest that the positive-direction continuum lags of the stream collision peak in TDEs are unlikely caused by the diffuse continuum as AGNs.

\begin{deluxetable*}{lcccccccc}
\tablecaption{{Properties of 7 TDEs with RB bump} }
\tablehead{\colhead{Name} & \colhead{$z$}  & \colhead{$t_{\rm start}$}  & \colhead{$t_{\rm peak}$} & \colhead{$t_{\rm duration}$} & \colhead{Log $M_{\rm BH}/M_{\odot}$}& \colhead{BH method} & \colhead{Ref}\\
&&(day)&(day)&(day)&&}
\startdata
         BH mass $\leq 10^{8}$ $M_{\odot}$:\\
          AT 2021uqv& 0.106 & 70 & 240 & 270 & 6.3$^{+0.4}_{-0.4}$&$M_{\rm BH} - \sigma_{*}$& \citep{Yao23} \\
          AT 2019avd & 0.028 &210&600&600 &6.3$^{+0.3}_{-0.3}$&single-epoch spec.&\citep{Malyali21,Wang23}\\
          AT 2020ocn & 0.070 & 40 & 60& 50 & 6.5$^{+0.7}_{-0.7}$&$M_{\rm BH} - \sigma_{*}$&\citep{Hammerstein23}\\
          AT 2019baf & 0.089 & 200 & 250 & 150 & 6.8$^{+0.3}_{-0.3}$&$M_{\rm BH} - M_{\rm gal}$& \citep{Yao23}\\
          AT 2018fyk & 0.059 & 30 & 70 & 550 &  7.7$^{+0.4}_{-0.4}$&$M_{\rm BH} - \sigma_{*}$ & \citep{Wevers23}\\ 
          \hline
        BH mass $> 10^{8}$ $M_{\odot}$:\\
          AT 2020acka& 0.338 & 80 & 210 & 600 &  8.2$^{+0.4}_{-0.4}$&$M_{\rm BH} - \sigma_{*}$& \citep{Yao23}\\
          ASASSN-15lh & 0.233 & 60 & 120 & $>$300 & 8.5$^{+0.5}_{-0.5}$& $M_{\rm BH}-L$& \citep{Leloudas16}\\
\enddata
\tablecomments{Redshift and BH mass are drawn from previous works listed in the last column. $t_{\rm start}$ is the start point of the RB bump, defined as the lowest point between the first and second peaks in Figure \ref{fig:RB_sources}. $t_{\rm peak}$ is defined as the highest point on the RB bump. $t_{\rm duration}$ is the duration time of RB bump, roughly estimated based on the convex region, although light curves are not complete. We found no potential correlations between black mass and different timescales. }
\label{tab:RB_infor}
\end{deluxetable*}

\begin{deluxetable*}{lcc}\label{tab:RB_prior}
\tablecaption{Priors for light-curve detrending using MCMC analysis.}
\tablehead{\colhead{Parameter} & \colhead{Description} & \colhead{Prior} }
\startdata
log $L_{\rm peak}$ & Peak luminosity  & [log ($L_{\rm max}/2$), log ($2L_{\rm max}$)]\\
$t_{\rm peak}$ & Peak time & [$-5$, 5] days\\
$t_{\rm 0}$ & Power-law normalization & [0, 3] days\\
$p$ & Power-law index & [$-5$, 0]\\
log $\tau$ & Exponential decay time & [0, 3] days\\
\enddata
\end{deluxetable*}

\begin{deluxetable*}{lcccccc}\label{tab:RB_detrending}
\tablecaption{Light curve detrending properties for 7 additional TDEs with RB bumps}
\tablehead{\colhead{Name} & \colhead{band} & \colhead{log $L_{\rm peak}$} & 
\colhead{$t_{\rm peak}$} & \colhead{p / log $\tau$} & \colhead{log $t_{0}$} &  \colhead{$\chi_{\rm \nu} ^{2}$}}
\startdata
\hline\hline
Name & band & log $L_{\rm peak}$ & $t_{\rm peak}$ & p / log $\tau$ & log $t_{0}$ &$\chi_{\rm \nu} ^{2}$\\
\hline
AT2021uqv
&g&$43.28_{-0.05}^{+0.06}$&$-0.28_{-3.36}^{+3.71}$&$-1.08_{-0.39}^{+0.20}$&$1.49_{-0.18}^{+0.21}$&1.79 \\
(PL)&r&$43.06_{-0.04}^{+0.04}$&$-0.43_{-3.24}^{+3.69}$&$-1.50_{-0.77}^{+0.42}$&$1.76_{-0.23}^{+0.24}$&2.32 \\
\hline
AT2019baf
&g&$43.54_{-0.04}^{+0.04}$&$0.10_{-3.42}^{+3.36}$&$-3.68_{-0.33}^{+0.30}$&$2.16_{-0.06}^{+0.06}$&1.42 \\
(PL)&r&$43.48_{-0.05}^{+0.05}$&$-0.10_{-3.40}^{+3.45}$&$-2.98_{-0.27}^{+0.23}$&$1.94_{-0.06}^{+0.06}$&1.92 \\
&i&$43.46_{-0.15}^{+0.28}$&$0.04_{-3.59}^{+3.28}$&$-2.10_{-1.07}^{+0.56}$&$1.66_{-0.45}^{+0.36}$&1.1 \\
\hline
AT2019avd
&g&$43.26_{-0.15}^{+0.24}$&$-1.76_{-2.28}^{+3.35}$&$-0.83_{-0.11}^{+0.09}$&$0.83_{-0.41}^{+0.27}$&0.58 \\
(PL)&r&$43.13_{-0.08}^{+0.15}$&$0.05_{-3.56}^{+4.42}$&$-1.01_{-0.46}^{+0.29}$&$1.40_{-0.49}^{+0.32}$&0.77 \\
\hline
\hline
AT2018fyk
&UVW2&$44.29_{-0.09}^{+0.09}$&$4.85_{-6.66}^{+6.63}$&$1.51_{-0.02}^{+0.02}$& &1.92\\
(Exp.)&UVM2&$44.12_{-0.07}^{+0.07}$&$4.71_{-6.61}^{+6.69}$&$1.60_{-0.03}^{+0.03}$& &1.73\\
&UVW1&$44.04_{-0.08}^{+0.08}$&$4.83_{-6.62}^{+6.63}$&$1.58_{-0.03}^{+0.03}$&&0.96\\
&U&$43.78_{-0.08}^{+0.09}$&$4.81_{-6.67}^{+6.60}$&$1.53_{-0.04}^{+0.04}$& &1.40\\
\hline
ASASSN-15lh
&UVW2&$45.83_{-0.02}^{+0.03}$&$1.92_{-1.70}^{+1.87}$&$2.97_{-0.04}^{+0.02}$& &2.23\\
(Exp.)&UVM2&$44.12_{-0.07}^{+0.07}$&$4.71_{-6.61}^{+6.69}$&$1.60_{-0.03}^{+0.03}$& &1.73\\
&UVW1&$44.04_{-0.08}^{+0.08}$&$4.83_{-6.62}^{+6.63}$&$1.58_{-0.03}^{+0.03}$& &0.96\\
&U&$43.78_{-0.08}^{+0.09}$&$4.81_{-6.67}^{+6.60}$&$1.53_{-0.04}^{+0.04}$&&1.40\\
&B&$45.04_{-0.04}^{+0.03}$&$-0.19_{-3.32}^{+3.55}$&$1.65_{-0.02}^{+0.02}$&&1.39\\
&g&$44.95_{-0.04}^{+0.04}$&$-0.05_{-3.45}^{+3.49}$&$1.66_{-0.04}^{+0.04}$&&0.10 \\
&V&$44.85_{-0.03}^{+0.03}$&$-0.04_{-3.38}^{+3.40}$&$1.68_{-0.02}^{+0.02}$&&1.06\\
&r&$44.62_{-0.03}^{+0.04}$&$-0.16_{-3.29}^{+3.54}$&$1.76_{-0.05}^{+0.06}$&&0.18\\
&i&$44.32_{-0.03}^{+0.03}$&$-0.26_{-3.29}^{+3.49}$&$1.88_{-0.06}^{+0.07}$&&0.41\\
\hline
AT2020acka
&g&$44.77_{-0.04}^{+0.03}$&$-0.08_{-3.29}^{+3.45}$&$1.63_{-0.01}^{+0.01}$&&4.06\\
(Exp.)&r&$44.45_{-0.03}^{+0.03}$&$-0.10_{-3.40}^{+3.44}$&$1.70_{-0.01}^{+0.01}$&&2.53\\
&i&$44.22_{-0.03}^{+0.03}$&$-0.14_{-3.39}^{+3.56}$&$1.70_{-0.03}^{+0.03}$&&1.28\\
\hline
AT2020ocn
&g&$42.56_{-0.02}^{+0.02}$&$-1.63_{-2.29}^{+2.30}$&$1.72_{-0.04}^{+0.04}$&&1.33\\
(Exp.)&r&$42.33_{-0.03}^{+0.03}$&$-1.62_{-2.27}^{+2.24}$&$1.60_{-0.04}^{+0.05}$&&0.91\\
&i&$42.09_{-0.04}^{+0.05}$&$-1.46_{-2.41}^{+2.34}$&$1.74_{-0.11}^{+0.14}$&&1.53\\
\hline
\enddata
\tablecomments{Note that $p$ and log $\tau$ describe the decay speed of the light curves in the power-law or exponential modeling. }
\end{deluxetable*}

\begin{deluxetable*}{lccccc}\label{tab:RB_lag}
\tablecaption{Lag measurements of the RB bump in four TDEs.}
\tablehead{\colhead{Name} & \colhead{Log $M_{\rm BH}/M_{\odot}$} & \colhead{X-ray vs. UV} & \colhead{X-ray vs. Opt} &\colhead{UV vs. Opt} & \colhead{Opt$-g$ vs. Opt$-r$}\\
& & (day)&(day) &(day)&(day)\\}
\startdata
ASASSN-15lh& 8.5 $\pm$ 0.5 &  &  & $15.0^{+3.8}_{-3.3} $  &   \\
AT 2019avd & 6.3 $\pm$ 0.3 & $-55.2^{+68.6}_{-28.9}$ &$-46.9^{+10.5}_{-16.5}$ & $19.3^{+13.8}_{-17.1} $ & 4.8$^{+2.9}_{-3.3} $   \\
AT 2020ocn & 6.5 $\pm$ 0.7 & $-42^{+12.1}_{-3.4}$ & $-42^{+12.1}_{-3.4}$ &  &  \\
AT 2018fyk (A) & 7.7 $\pm$ 0.4 & $9.2^{+6.1}_{-6.0}$ &  &   &    \\
AT 2018fyk (B) & 7.7 $\pm$ 0.4 & $-4.0^{+1.4}_{-1.4}$ &  &  &  \\
\enddata
\tablecomments{Note that stage A of AT 2018fyk is on the RB bump, while stage B is after the RB bump. Empty cell indicates no lag is detectable due to, e.g., low light-curve cadence, no significant variabilities, or too large lag uncertainties. }
\end{deluxetable*}

\end{document}